\documentclass[12pt,fleqn,a4paper]{article} % A4 size
\usepackage{epsfig,graphics,graphicx}
\usepackage{clshan-math,clshan-dm-dd}
\newcommand{\imageswitch} [2] {#2}
\def \lsim {\:\raisebox{-0.7ex}{$\stackrel{\textstyle<}{\sim}$}\:}
\def \gsim {\:\raisebox{-0.7ex}{$\stackrel{\textstyle>}{\sim}$}\:}
\begin{document}
\thispagestyle{empty}
\begin{flushright}
 March 2010
\end{flushright}
\begin{center}
{\large\bf
 Effects of Residue Background Events
 in Direct Dark Matter \\ \vspace{0.1cm} Detection Experiments on
 the Determination of the WIMP Mass} \\
\vspace*{0.7cm}
 {\sc Yu-Ting Chou}$^{1}$
  and {\sc Chung-Lin Shan}$^{2, 3}$ \\
\vspace{0.5cm}
 ${}^1$
 {\it Institute of Physics, National Chiao Tung University  \\
      No.~1001, University Road,
      Hsinchu City 30010, Taiwan, R.O.C.}                   \\
%\vspace{0.1cm}
 {\it E-mail:} {\tt yuting.py97g@nctu.edu.tw}               \\~\\
 ${}^2$
 {\it Department of Physics, National Cheng Kung University \\
      No.~1, University Road,
      Tainan City 70101, Taiwan, R.O.C.}                    \\
 {\it E-mail:} {\tt clshan@mail.ncku.edu.tw}                \\~\\
%\vspace{0.1cm}
%
 ${}^3$
 {\it Physics Division,
      National Center for Theoretical Sciences              \\
      No.~101, Sec.~2, Kuang-Fu Road,
      Hsinchu City 30013, Taiwan, R.O.C.}                   \\
%\vspace{0.1cm}
%
\end{center}
\vspace{1cm}
\begin{abstract}
 In the earlier work on the development of
 a model--independent data analysis method
 for determining the mass of
 Weakly Interacting Massive Particles (WIMPs)
 by using measured recoil energies
 from direct Dark Matter detection experiments directly,
 it was assumed that
 the analyzed data sets are background--free,
 i.e., all events are WIMP signals.
 In this article,
 as a more realistic study,
 we take into account
 a fraction of possible residue background events,
 which pass all discrimination criteria and
 then mix with other real WIMP--induced events
 in our data sets.
 Our simulations show that,
 for the determination of the WIMP mass,
 the maximal acceptable fraction of residue background events
 in the analyzed data sets
 of ${\cal O}(50)$ total events
 is $\sim$ 20\%,
 for background windows of
 the entire experimental possible energy ranges,
 or in low energy ranges;
 while,
 for background windows in relatively higher energy ranges,
 this maximal acceptable fraction of residue background events
 can not be larger than $\sim$ 10\%.
 For a WIMP mass of 100 GeV
 with 20\% background events
 in the windows of
 the entire experimental possible energy ranges,
 the reconstructed WIMP mass
 and the 1$\sigma$ statistical uncertainty
 are $\sim 97~{\rm GeV}\~^{+61\%}_{-35\%}$
 ($\sim 94~{\rm GeV}\~^{+55\%}_{-33\%}$
  for background--free data sets).
\end{abstract}
\clearpage
\section{Introduction}
 Currently,
 direct Dark Matter detection experiments
 searching for Weakly Interacting Massive Particles (WIMPs)
 are one of the promising methods
 for understanding the nature of Dark Matter
 and identifying them among new particles produced at colliders
 as well as reconstructing the (sub)structure of our Galactic halo
 \cite{Smith90, Lewin96, SUSYDM96, Bertone05}.
 In order to determine the mass of halo WIMPs
 {\em without} making any assumptions
 about their density near the Earth or
 their velocity distribution
 {\em nor} knowing their scattering cross section on nucleus,
 a model--independent method
 by combining two experimental data sets
 with two different target nuclei
 has been developed
 \cite{DMDDmchi-SUSY07, DMDDmchi}.
 This method builds on the earlier work
 on the reconstruction of the (moments of the)
 one--dimensional velocity distribution function of halo WIMPs,
 $f_1(v)$,
 by using data
 from direct detection experiments
 \cite{DMDDf1v}.

 In the analysis of reconstructing $f_1(v)$,
 the moments of the WIMP velocity distribution function
 can be determined from experimental data directly
 with an {\em unique} input information
 about the WIMP mass $\mchi$.
 Hence,
 one can simply require that
 the values of a given moment of $f_1(v)$
 determined by two experiments agree%
\footnote{
 Note that,
 as demonstrated and discussed
 in Ref.~\cite{DMDDmchi},
 this condition requires an algorithmic procedure
 for matching the maximal cut--off energies
 of the analyzed data sets.
}.
 This leads to a simple analytic expression
 for determining $\mchi$
 \cite{DMDDmchi-SUSY07, DMDDmchi},
 where each moment can in principle be used.
 Additionally,
 under the assumptions that
 the spin--independent (SI) WIMP--nucleus interaction
 dominates over the spin--dependent (SD) one
 and the SI WIMP coupling on protons
 is approximately the same as that on neutrons,
 a second analytic expression for determining $\mchi$
 has been derived \cite{DMDDmchi}.
 Finally,
 by combining the first estimators for different moments
 with each other and with the second estimator,
 one can yield the best--fit WIMP mass
 as well as minimize its statistical uncertainty.

 In the work on the development of
 the model--independent data analysis procedure
 for the determination of the WIMP mass,
 it was assumed that
 the analyzed data sets are background--free,
 i.e., all events are WIMP signals.
 Active background discrimination techniques
 should make this condition possible.
 For example,
 the ratio of the ionization to recoil energy,
 the so--called ``ionization yield'',
 used in the CDMS-II experiment
 provides an event--by--event rejection
 of electron recoil events
 to be better than $10^{-4}$ misidentification
 \cite{Ahmed09b}.
 By combining the ``phonon pulse timing parameter'',
 the rejection ability of
 the misidentified electron recoils
 (most of them are ``surface events''
  with sufficiently reduced ionization energies)
 can be improved to be $< 10^{-6}$ for electron recoils
 \cite{Ahmed09b}.
 Moreover,
 as demonstrated by the CRESST collaboration
 \cite{CRESST-bg}, % Lang09a, Schmaler09},
 by means of inserting a scintillating foil,
 which causes some additional scintillation light
 for events induced by $\alpha$-decay of $\rmXA{Po}{210}$
 and thus shifts the pulse shapes of these events
 faster than pulses induced by WIMP interactions in the crystal,
 the pulse shape discrimination (PSD) technique
 can then easily distinguish WIMP--induced nuclear recoils
 from those induced by backgrounds%
\footnote{
 For more details
 about background discrimination techniques and status
 in currently running and projected direct detection experiments,
 see e.g.,
 Refs.~\cite{Aprile09a,
             EDELWEISS-bg, % Broniatowski09,
             Lang09b} %, Armengaud09}.
}.

 However,
 as the most important issue in all underground experiments,
 the signal identification ability and
 possible residue background events
 which pass all discrimination criteria and
 then mix with other real WIMP--induced events in our data sets
 should also be considered.
 Therefore,
 in this article,
 as a more realistic study,
 we take into account
 different fractions of residue background events
 mixed in experimental data sets
 and want to study
 how well the model--independent method
 could reconstruct the input WIMP mass
 by using these ``impure'' data sets
 and how ``dirty'' these data sets could be
 to be still useful.

 The remainder of this article is organized as follows.
 In Sec.~2
 we review the recoil spectrum of elastic WIMP--nucleus scattering
 and introduce two kinds of background spectrum
 used in our simulations.
 In Sec.~3
 we first review briefly
 the model--independent method
 for the determination of the WIMP mass.
 Then we show numerical results of the reconstructed WIMP mass
 by using mixed data sets
 with different fractions of residue background events
 based on Monte Carlo simulations.
 We conclude in Sec.~4.
 Some technical details will be given in an appendix.
\section{Signal and background spectra}
 In this section
 we first review
 the recoil spectrum of elastic WIMP--nucleus scattering.
 Then we introduce two forms of background spectrum
 which will be used in our simulations.
 Some numerical results of the measured energy spectrum
 superposed by the WIMP scattering and background spectra
 will also be discussed.
\subsection{Elastic WIMP--nucleus scattering spectrum}
 The basic expression for the differential event rate
 for elastic WIMP--nucleus scattering is given by \cite{SUSYDM96}:
\beq
   \dRdQ
 = \calA \FQ \int_{\vmin}^{\vmax} \bfrac{f_1(v)}{v} dv
\~.
\label{eqn:dRdQ}
\eeq
 Here $R$ is the direct detection event rate,
 i.e., the number of events
 per unit time and unit mass of detector material,
 $Q$ is the energy deposited in the detector,
 $F(Q)$ is the elastic nuclear form factor,
 $f_1(v)$ is the one--dimensional velocity distribution function
 of the WIMPs impinging on the detector,
 $v$ is the absolute value of the WIMP velocity
 in the laboratory frame.
 The constant coefficient $\calA$ is defined as
\beq
        \calA
 \equiv \frac{\rho_0 \sigma_0}{2 \mchi \mrN^2}
\~,
\label{eqn:calA}
\eeq
 where $\rho_0$ is the WIMP density near the Earth
 and $\sigma_0$ is the total cross section
 ignoring the form factor suppression.
 The reduced mass $\mrN$ is defined by
\beq
        \mrN
 \equiv \frac{\mchi \mN}{\mchi + \mN}
\~,
\label{eqn:mrN}
\eeq
 where $\mchi$ is the WIMP mass and
 $\mN$ that of the target nucleus.
 Finally,
 $\vmin$ is the minimal incoming velocity of incident WIMPs
 that can deposit the energy $Q$ in the detector:
\beq
   \vmin
 = \alpha \sqrt{Q}
\label{eqn:vmin}
\eeq
 with the transformation constant
\beq
        \alpha
 \equiv \sfrac{\mN}{2 \mrN^2}
\~,
\label{eqn:alpha}
\eeq
 and $\vmax$ is the maximal WIMP velocity
 in the Earth's reference frame,
 which is related to
 the escape velocity from our Galaxy
 at the position of the Solar system,
 $\vesc~\gsim~600$ km/s.
 Note that,
 as will be shown below,
 the Earth's velocity relative to the Galactic halo
 is time--dependent,
 and considering the random motion of WIMPs in the Galaxy,
 the relation between
 the one--dimensional cut--off $\vmax$ and
 the three--dimensional one $\vesc$
 is thus rather complicated.
 Nevertheless,
 it is unlike to affect the event rate
 as well as the results shown in this article
 significantly.
 In the literature,
 for simplicity and practical uses,
 $\vmax$ is often set as $\infty$
 (e.g., \cite{Bernal08, Green-mchi08, Cerdeno10}).
\subsubsection{One--dimensional WIMP velocity distribution function}
 The simplest semi--realistic model halo is
 a spherical isothermal Maxwellian halo.
 More realistically,
 one has to take into account
 the orbital motion of the Solar system around the Galaxy
 as well as that of the Earth around the Sun.
 The one--dimensional velocity distribution function of
 this shifted Maxwellian halo
 can be expressed as
 \cite{Lewin96, SUSYDM96, DMDDf1v}
\beq
   f_{1, \sh}(v)
 = \frac{1}{\sqrt{\pi}} \afrac{v}{\ve v_0}
   \bbigg{ e^{-(v - \ve)^2 / v_0^2} - e^{-(v + \ve)^2 / v_0^2} }
\~.
\label{eqn:f1v_sh}
\eeq
 Here $v_0 \simeq 220~{\rm km/s}$
 is the orbital velocity of the Sun
 in the Galactic frame,
 and $\ve$ is the Earth's velocity in the Galactic frame
 \cite{Freese88, SUSYDM96, Bertone05}:
\beq
   v_{\rm e}(t)
 = v_0 \bbrac{1.05 + 0.07 \cos\afrac{2 \pi (t - t_{\rm p})}{1~{\rm yr}}}
\~;
\label{eqn:ve}
\eeq
 $t_{\rm p} \simeq$ June 2nd is the date
 on which the velocity of the Earth
 relative to the WIMP halo is maximal.
 Substituting Eq.~(\ref{eqn:f1v_sh}) into Eq.~(\ref{eqn:dRdQ}),
 an analytic form of the integral over
 the velocity distribution function
 can be given as \cite{DMDDmchi-NJP}
\beqn
        \int_{\vmin}^{\vmax} \bfrac{f_{1, \sh}(v)}{v} dv
 \=     \frac{1}{2 \ve}
        \cBiggl{ \bbigg{ \erf{\T\afrac{\alpha \sqrt{Q} + \ve}{v_0}}
                        -\erf{\T\afrac{\alpha \sqrt{Q} - \ve}{v_0}} } }
        \non\\
 \conti ~~~~~~~~~~~~~~~~ %16
        \cBiggr{-\bbigg{ \erf{\T\afrac{\vmax           + \ve}{v_0}}
                        -\erf{\T\afrac{\vmax           - \ve}{v_0}} } }
\~.
\label{eqn:int_f1v_sh}
\eeqn
 Here $\erf(x)$ is the error function, defined as
\beqN
   \erf(x)
 = \frac{2}{\sqrt{\pi}} \int_0^{x} e^{-t^2} dt
\~.
\eeqN

 On the other hand,
 for practical, numerical uses,
 an approximate form of the integral over $f_1(v)$
 was introduced as \cite{Lewin96}
\beq
   \int_{\vmin}^{\infty} \bfrac{f_1(v)}{v} dv
 = c_0 \afrac{2}{\sqrt{\pi} v_0} e^{-\alpha^2 Q/c_1 v_0^2}
\~,
\label{eqn:int_f1v_cc}
\eeq
 where $c_0$ and $c_1$ are two fitting parameters of order unity.
 Not surprisingly,
 their values depend on
 the Galactic orbital and escape velocities,
 the target nucleus,
 the threshold energy of the experiment,
 as well as on the mass of incident WIMPs.
 Note that,
 the characteristic energy $Q_{\rm ch} \equiv c_1 v_0^2 / \alpha^2$
%
%\beq
%        Q_{\rm ch}
% \equiv \frac{c_1 v_0^2}{\alpha^2}
%\label{eqn:Q_ch}
%\eeq
%
 and thus the shape of the recoil spectrum
 depend highly on the WIMP mass:
 for light WIMPs ($\mchi \ll m_{\rm N}$),
 $Q_{\rm ch} \propto \mchi^2$ and
 the recoil spectrum drops sharply with increasing recoil energy,
 while for heavy WIMPs ($\mchi \gg m_{\rm N}$),
 $Q_{\rm ch} \sim$ const.~and the spectrum becomes flatter.
\subsubsection{Spin--independent WIMP--nucleus cross section}
 In most theoretical models,
 the spin--independent (SI) WIMP--nucleus interaction
 with an atomic mass number $A~\gsim~30$ dominates over
 the spin--dependent (SD) one
 \cite{SUSYDM96, Bertone05}.
 Additionally,
 for the lightest supersymmetric neutralino
 which is perhaps the best motivated WIMP candidate
 \cite{SUSYDM96, Bertone05},
 and for all WIMPs which interact primarily through Higgs exchange,
 the SI scalar coupling is approximately the same
 on both protons p and neutrons n,
 the ``pointlike'' cross section $\sigma_0$ in Eq.~(\ref{eqn:calA})
 can thus be written as
\beq
   \sigma_0
 = A^2 \afrac{m_{\rm r, N}}{m_{\rm r, p}}^2 \sigmapSI
\~,
\label{eqn:sigma0SI}
\eeq
 where
\beq
   \sigmapSI
 = \afrac{4}{\pi} m_{\rm r, p}^2 |f_{\rm p}|^2
%\~,
\label{eqn:sigmapSI}
\eeq
 is the SI WIMP--proton cross section,
 $f_{\rm p}$ is the effective $\chi \chi {\rm p p}$ four--point coupling,
 $A$ is the atomic mass number of the target nucleus,
 and $m_{\rm r, p}$ is the reduced mass
 of the WIMP mass $\mchi$
 and the proton mass $m_{\rm p}$.

 For the SI WIMP--nucleus cross section,
 an analytic form for the elastic nuclear form factor,
 inspired by the Woods--Saxon nuclear density profile,
 has been suggested by Engel as
 \cite{Engel91, SUSYDM96, Bertone05}%
\footnote{
 Other commonly used analytic forms
 for the nuclear form factor
 for the SI WIMP--nucleus cross section
 can be found in Ref.~\cite{DMDDmchi-NJP}.
}
\beq
   F_{\rm WS}^2(Q)
 = \bfrac{3 j_1(q R_1)}{q R_1}^2 e^{-(q s)^2}
\~.
\label{eqn:FQ_SI_WS}
\eeq
 Here $j_1(x)$ is a spherical Bessel function,
 $\D q = \sqrt{2 m_{\rm N} Q}$ is the transferred 3-momentum,
 given as a function of
 the recoil energy transferred
 from the incident WIMP to the target nucleus, $Q$,
 and the mass of the target nucleus, $\mN$;
%
%\beq
%   q
% = \sqrt{2 m_{\rm N} Q}
%\~,
%\label{eqn:qq}
%\eeq
%
 $R_1 = \sqrt{R_A^2 - 5 s^2}$ is the effective nuclear radius % defined by
%
%\beq
%   R_1
% = \sqrt{R_A^2 - 5 s^2}
%\label{eqn:R1}
%\eeq
%
 with $R_A \simeq 1.2 \~ A^{1/3}~{\rm fm}$
%
%\beq
%        R_A
% \simeq 1.2 \~ A^{1/3}~{\rm fm},
%\label{eqn:RA}
%\eeq
%
 and the nuclear skin thickness $s \simeq 1~{\rm fm}$.
%
%\beq
%        s
% \simeq 1~{\rm fm}
%\~.
%\label{eqn:ss}
%\eeq
%
%
\subsection{Background spectrum}
 For our simulations with residue background events,
 two forms of background spectrum
 are considered.
 The simplest choice for the background spectrum
 is the constant spectrum:
\beq
   \aDd{R}{Q}_{\rm bg, const}
 = 1
\~.
\label{eqn:dRdQ_bg_const}
\eeq
 More realistically,
 inspired by Ref.~\cite{Green-mchi08},
 we introduce a {\em target--dependent exponential}
 spectrum given by
\beq
   \aDd{R}{Q}_{\rm bg, ex}
 = \exp\abrac{-\frac{Q /{\rm keV}}{A^{0.6}}}
\~.
\label{eqn:dRdQ_bg_ex}
\eeq
 Here $Q$ is the recoil energy,
 $A$ is the atomic mass number of the target nucleus.
 The power index of $A$, 0.6, is an empirical constant,
 which has been chosen so that
 the exponential background spectrum is
 somehow {\em similar to},
 but still {\em different from}
 the expected recoil spectrum of the target nuclei;
 otherwise,
 there is in practice no difference between
 the WIMP scattering and background spectra.
 Note that,
 among different possible choices
 (e.g., the exponential form
  used in Ref.~\cite{Green-mchi08}),
 we use in our simulations the atomic mass number $A$
 as the simplest, unique characteristic parameter
 in the general analytic form (\ref{eqn:dRdQ_bg_ex})
 for defining the residue background spectrum
 for {\em different} target nuclei.
 However,
 it does {\em not} mean that
 the (superposition of the real) background spectra
 would depend simply/primarily on $A$ or
 on the mass of the target nucleus, $\mN$.
 In other words,
 it is practically equivalent to
 use expression (\ref{eqn:dRdQ_bg_ex})
 or $(dR / dQ)_{\rm bg, ex} = e^{-Q / 13.5~{\rm keV}}$ directly
 for a $\rmXA{Ge}{76}$ target.

 Note also that,
 firstly,
 two forms of background spectrum
 given in Eqs.~(\ref{eqn:dRdQ_bg_const}) and (\ref{eqn:dRdQ_bg_ex})
 are rather naive;
 however,
 since we consider here
 {\em only a few residue} background events
 induced by perhaps {\em two or more} different sources,
 pass all discrimination criteria,
 and then mix with other WIMP--induced events
 in our data sets of ${\cal O}(50)$ {\em total} events,
 exact forms of different background spectra
 are actually not very important and
 these two spectra,
 in particular,
 the exponential one,
 should practically not be unrealistic%
\footnote{
 Other (more realistic) forms for background spectrum
 (perhaps also for some specified targets/experiments)
 can be tested on the \amidas\ website
 \cite{AMIDAS-web, AMIDAS-eprints}. % AMIDAS-SUSY09, AMIDAS-f1vFQ}.
}.
 Secondly,
 for using the maximum likelihood analysis
 to determine the WIMP mass,
 as described in Refs.~\cite{Green-mchi07, Green-mchi08, Bernal08},
 a prior knowledge about the WIMP scattering spectrum
 and eventually about the background spectrum
 is essential \cite{Green-mchi08}.
 In contrast,
 as demonstrated in Ref.~\cite{DMDDmchi}
 and will be reviewed in the next section,
 the model--independent data analysis procedure
 requires only measured recoil energies
 (induced mostly by WIMPs and
  occasionally by background sources)
 from two experimental data sets
 with different target nuclei.
 Therefore,
 for applying this method to future real data
 from direct detection experiments,
 the prior knowledge about (different) background source(s)
 is {\em not required at all}.
\subsection{Measured energy spectrum}
\begin{figure}[p!]
\begin{center}
\vspace{-0.75cm}
\imageswitch{
\begin{picture}(16.5,20.5)
\put(0  ,14  ){\framebox(8,6.5){dRdQ-bg-ex-Ge-000-100-20-010}}
\put(8.5,14  ){\framebox(8,6.5){dRdQ-bg-ex-Ge-000-100-20-025}}
\put(0  , 7  ){\framebox(8,6.5){dRdQ-bg-ex-Ge-000-100-20-050}}
\put(8.5, 7  ){\framebox(8,6.5){dRdQ-bg-ex-Ge-000-100-20-100}}
\put(0  , 0  ){\framebox(8,6.5){dRdQ-bg-ex-Ge-000-100-20-250}}
\put(8.5, 0  ){\framebox(8,6.5){dRdQ-bg-ex-Ge-000-100-20-500}}
\end{picture}}
{\hspace*{-1.6cm}
 \includegraphics[width=9.8cm]{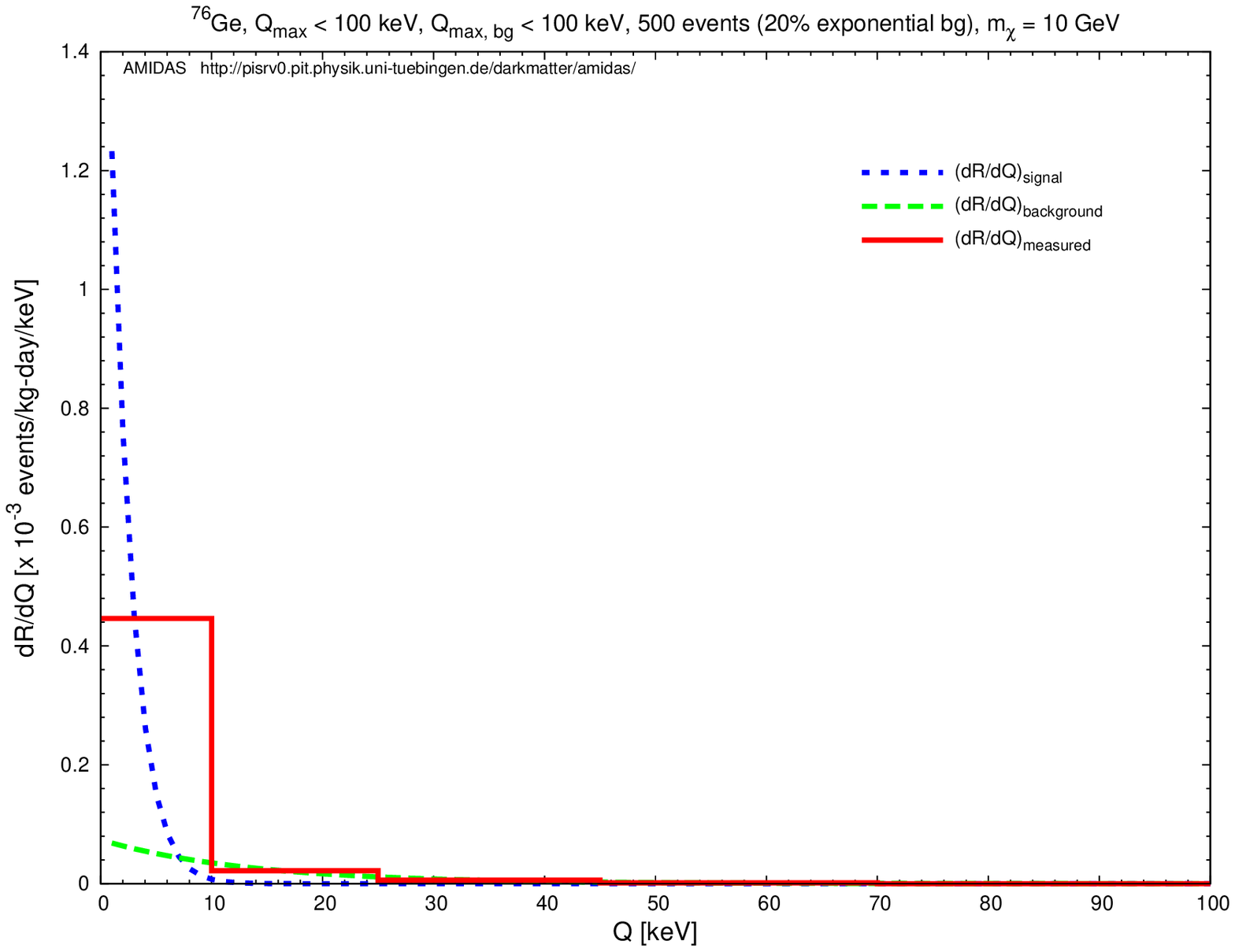} \hspace{-1.1cm}
 \includegraphics[width=9.8cm]{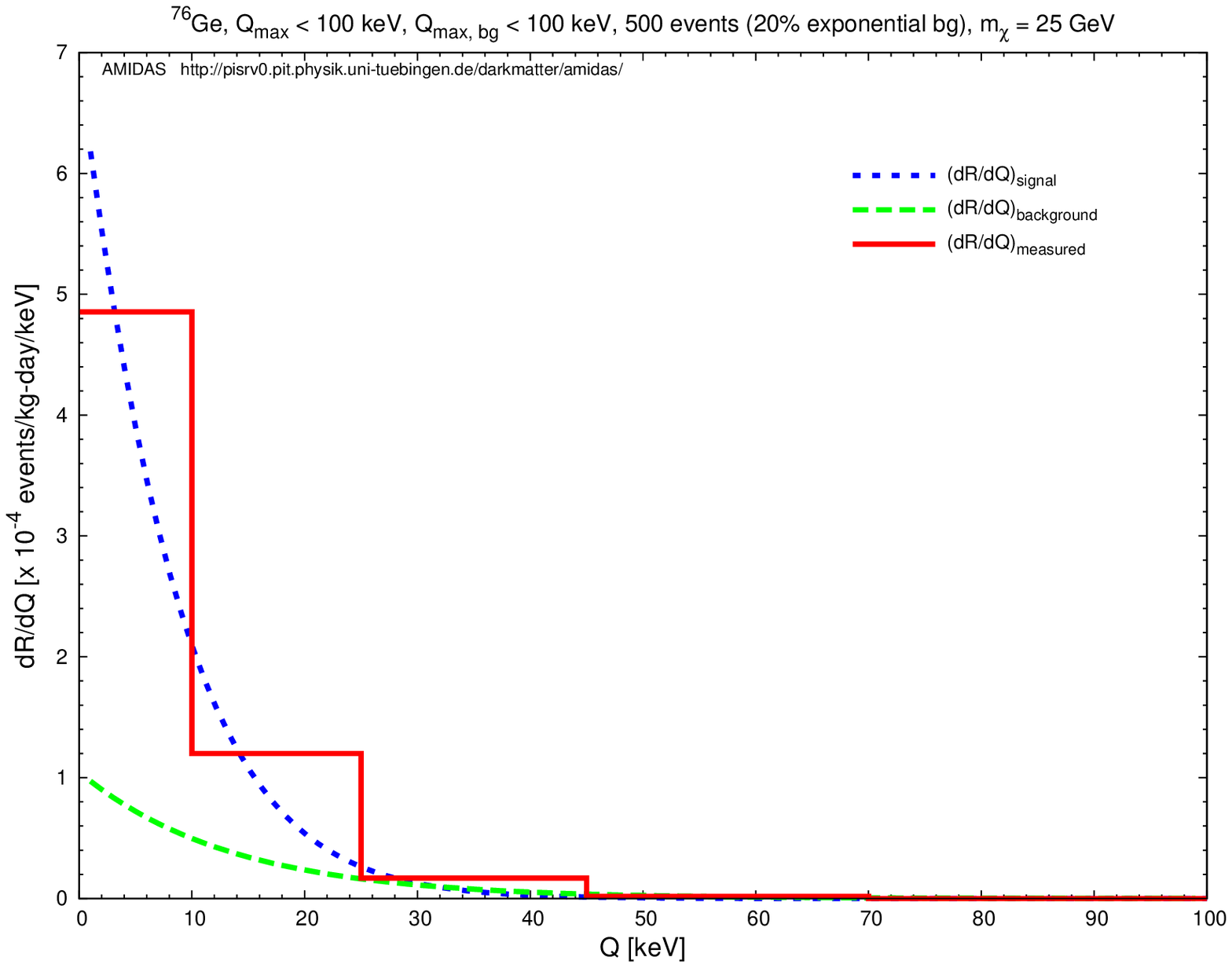} \hspace*{-1.6cm} \\
% \vspace{1cm}
 \hspace*{-1.6cm}
 \includegraphics[width=9.8cm]{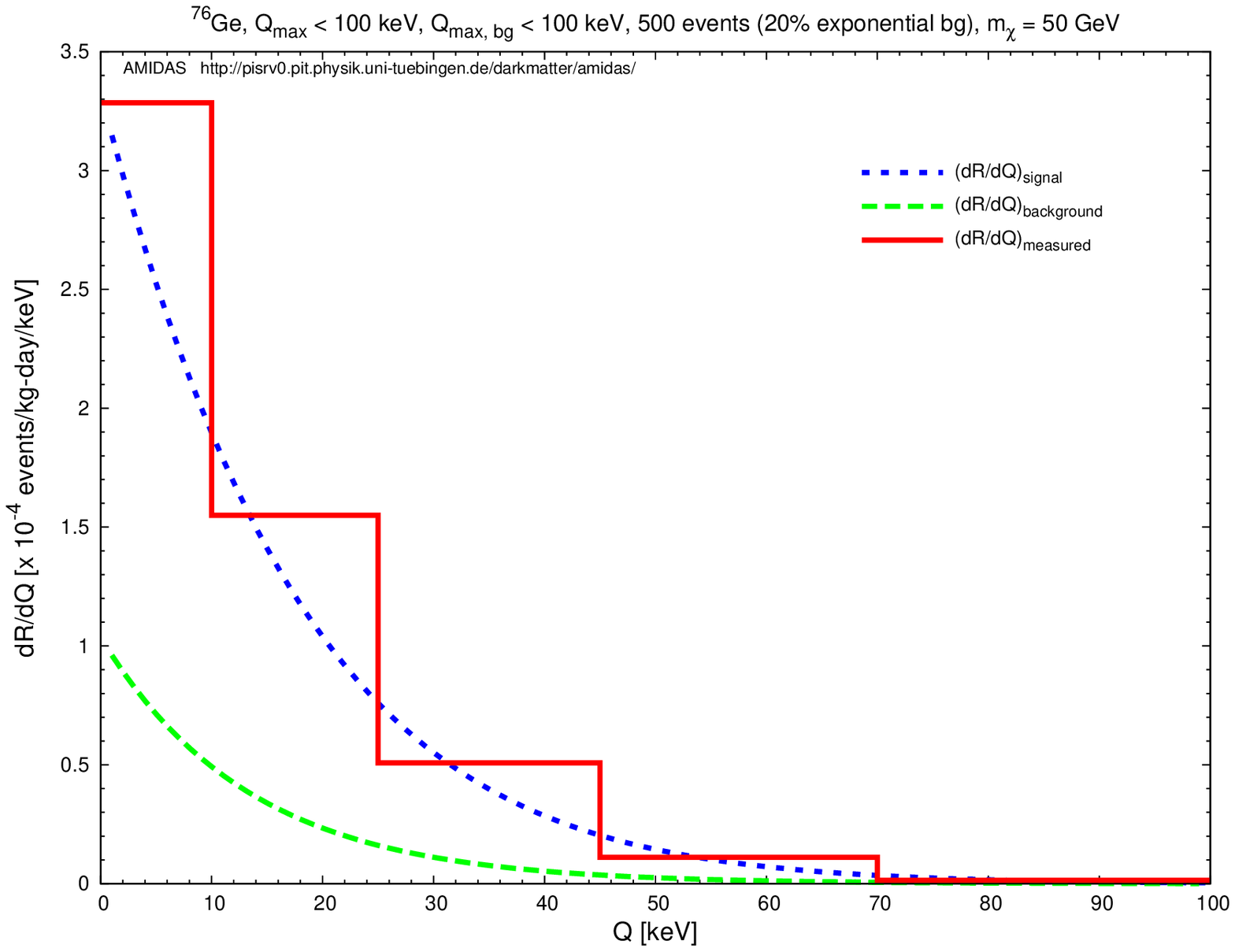} \hspace{-1.1cm}
 \includegraphics[width=9.8cm]{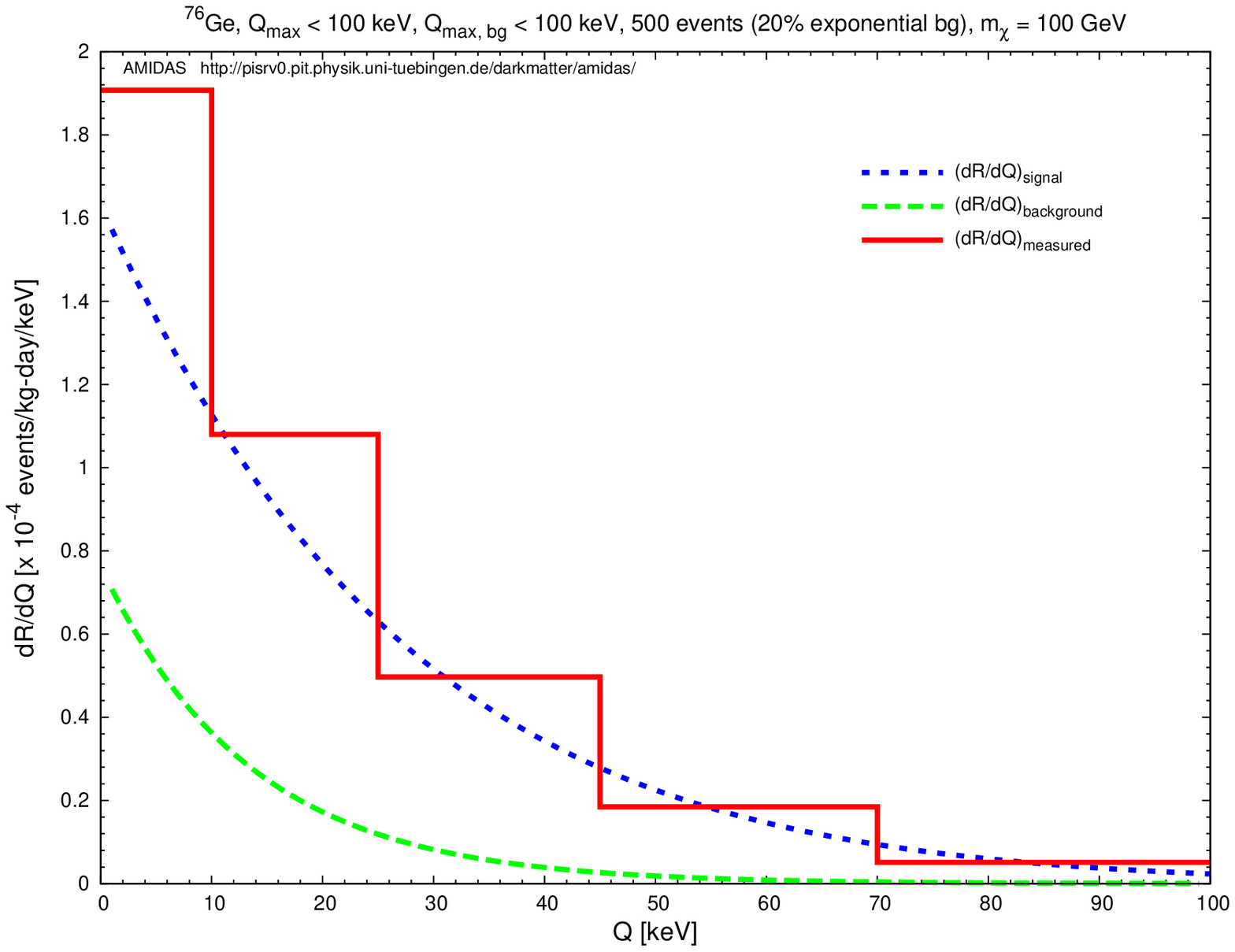} \hspace*{-1.6cm} \\
% \vspace{1cm}
 \hspace*{-1.6cm}
 \includegraphics[width=9.8cm]{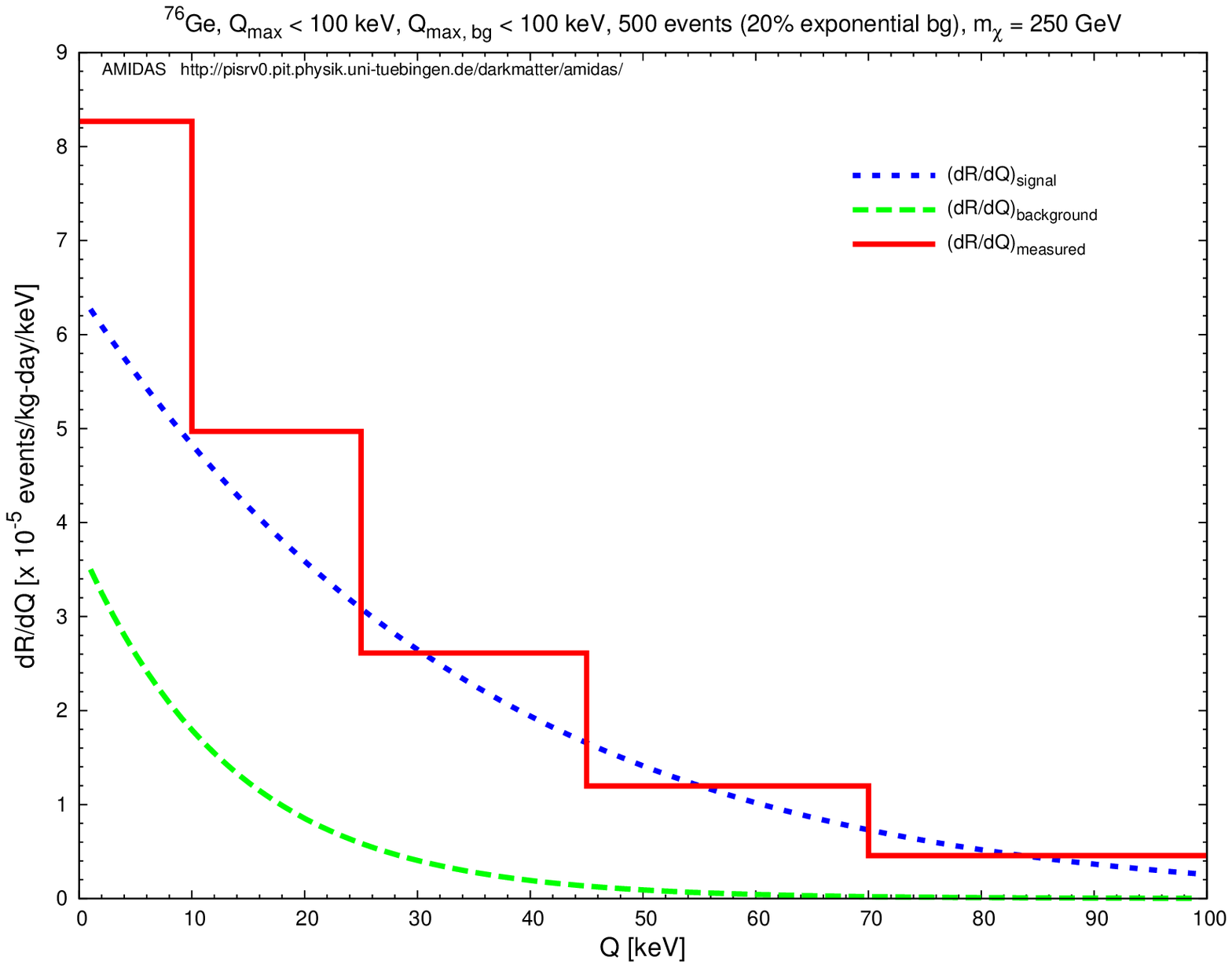} \hspace{-1.1cm}
 \includegraphics[width=9.8cm]{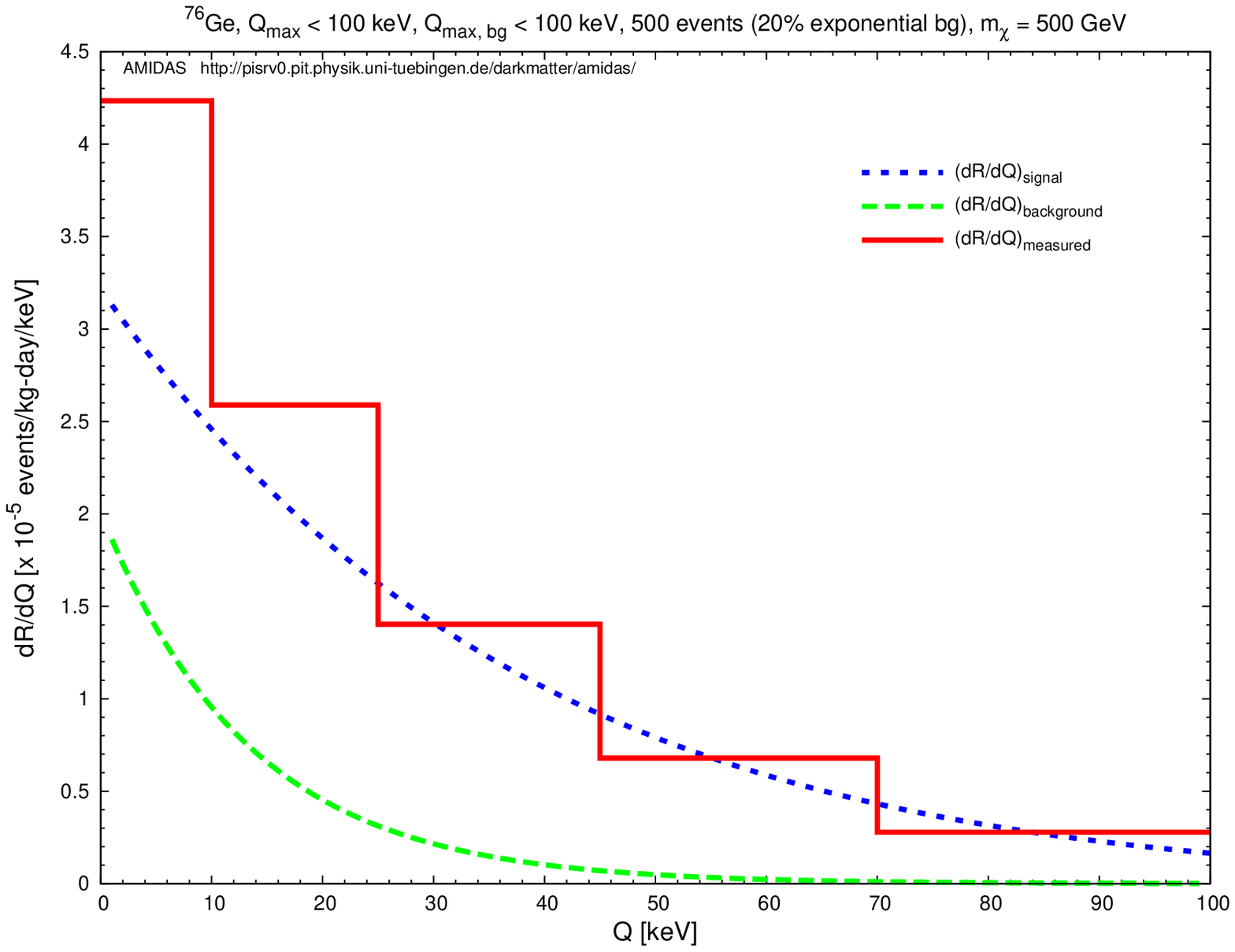} \hspace*{-1.6cm} \\
}
\vspace{-0.5cm}
\end{center}
\caption{
 Measured energy spectra (solid red histograms)
 for a $\rmXA{Ge}{76}$ target
 with six different WIMP masses:
 10, 25, 50, 100, 250, and 500 GeV.
 The dotted blue curves are
 the elastic WIMP--nucleus scattering spectra
 for the shifted Maxwellian velocity distribution
 and the Woods--Saxon elastic nuclear form factor;
 whereas
 the dashed green curves are
 the exponential background spectra
 normalized to fit to the chosen background ratio,
 which has been set as 20\% here.
 The experimental threshold energy
 has been assumed to be negligible
 and the maximal cut--off energy
 is set as 100 keV.
 The background windows
 have been assumed to be the same as
 the experimental possible energy ranges.
 5,000 experiments with 500 total events on average
 in each experiment have been simulated.
 See the text for further details.
}
\label{fig:dRdQ-bg-ex-Ge-000-100-20}
\end{figure}
\begin{figure}[p!]
\begin{center}
\imageswitch{
\begin{picture}(16.5,20.5)
\put(0  ,14  ){\framebox(8,6.5){dRdQ-bg-const-Ge-000-100-20-010}}
\put(8.5,14  ){\framebox(8,6.5){dRdQ-bg-const-Ge-000-100-20-025}}
\put(0  , 7  ){\framebox(8,6.5){dRdQ-bg-const-Ge-000-100-20-050}}
\put(8.5, 7  ){\framebox(8,6.5){dRdQ-bg-const-Ge-000-100-20-100}}
\put(0  , 0  ){\framebox(8,6.5){dRdQ-bg-const-Ge-000-100-20-250}}
\put(8.5, 0  ){\framebox(8,6.5){dRdQ-bg-const-Ge-000-100-20-500}}
\end{picture}}
{\hspace*{-1.6cm}
 \includegraphics[width=9.8cm]{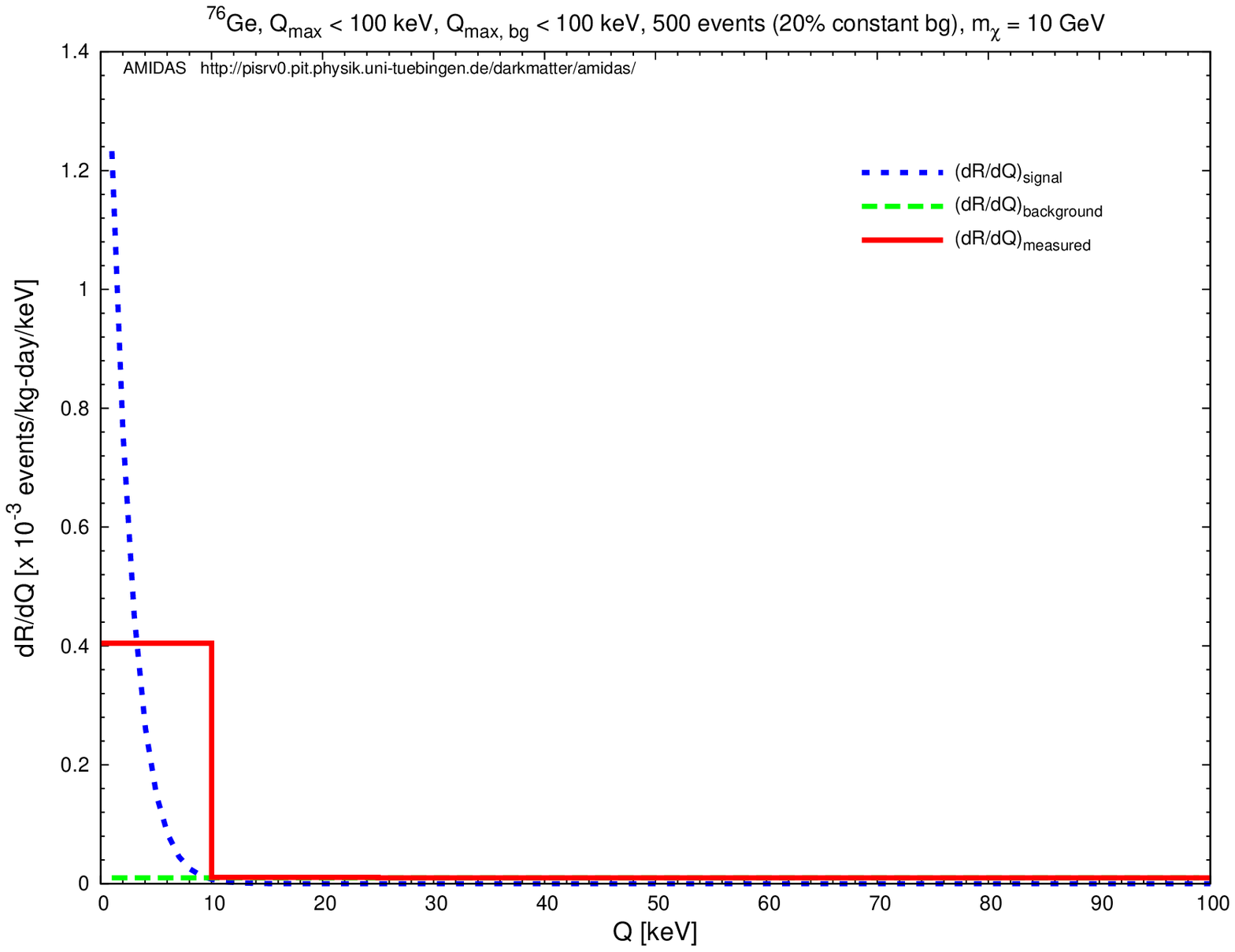} \hspace{-1.1cm}
 \includegraphics[width=9.8cm]{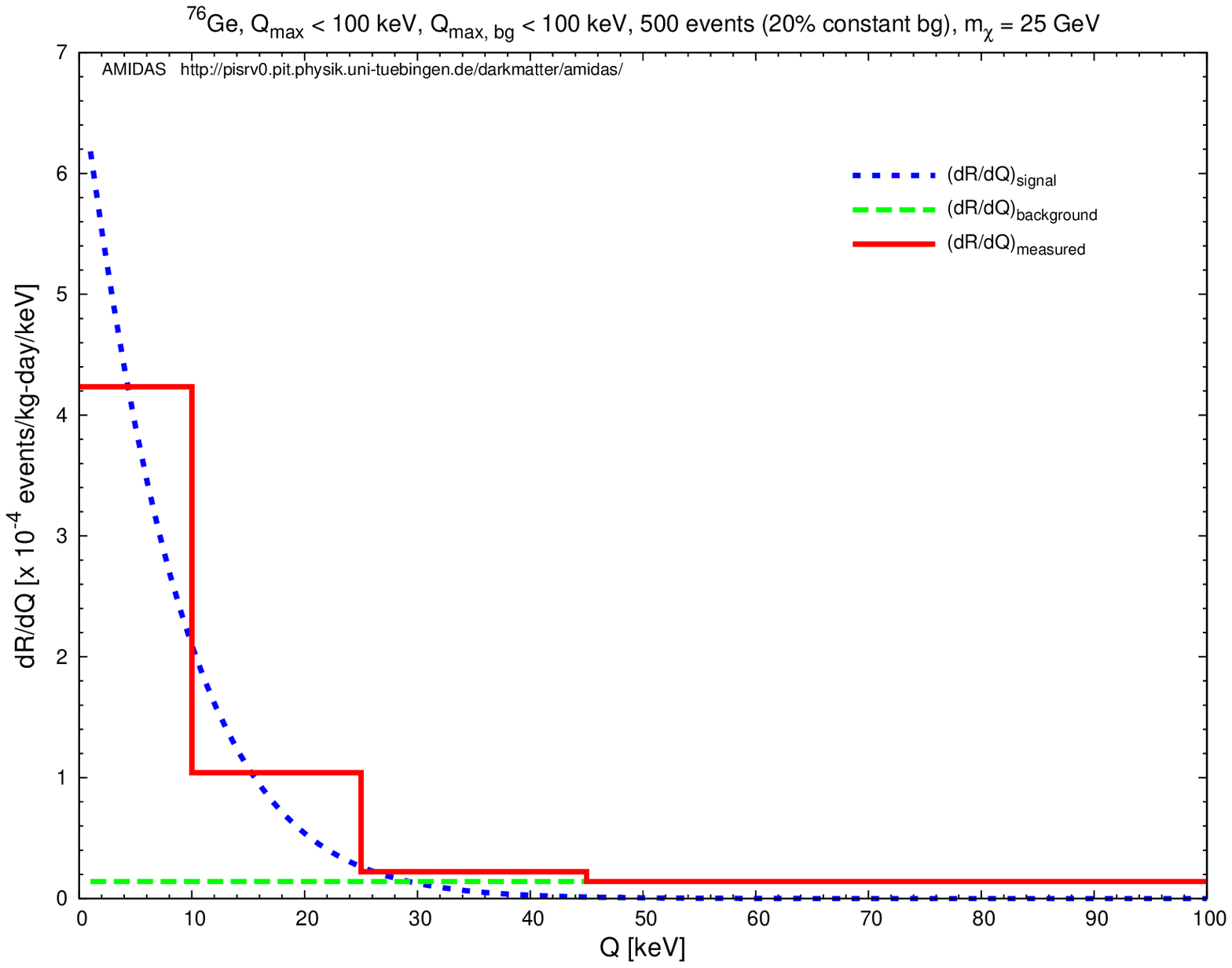} \hspace*{-1.6cm} \\
 \vspace{1cm}
 \hspace*{-1.6cm}
 \includegraphics[width=9.8cm]{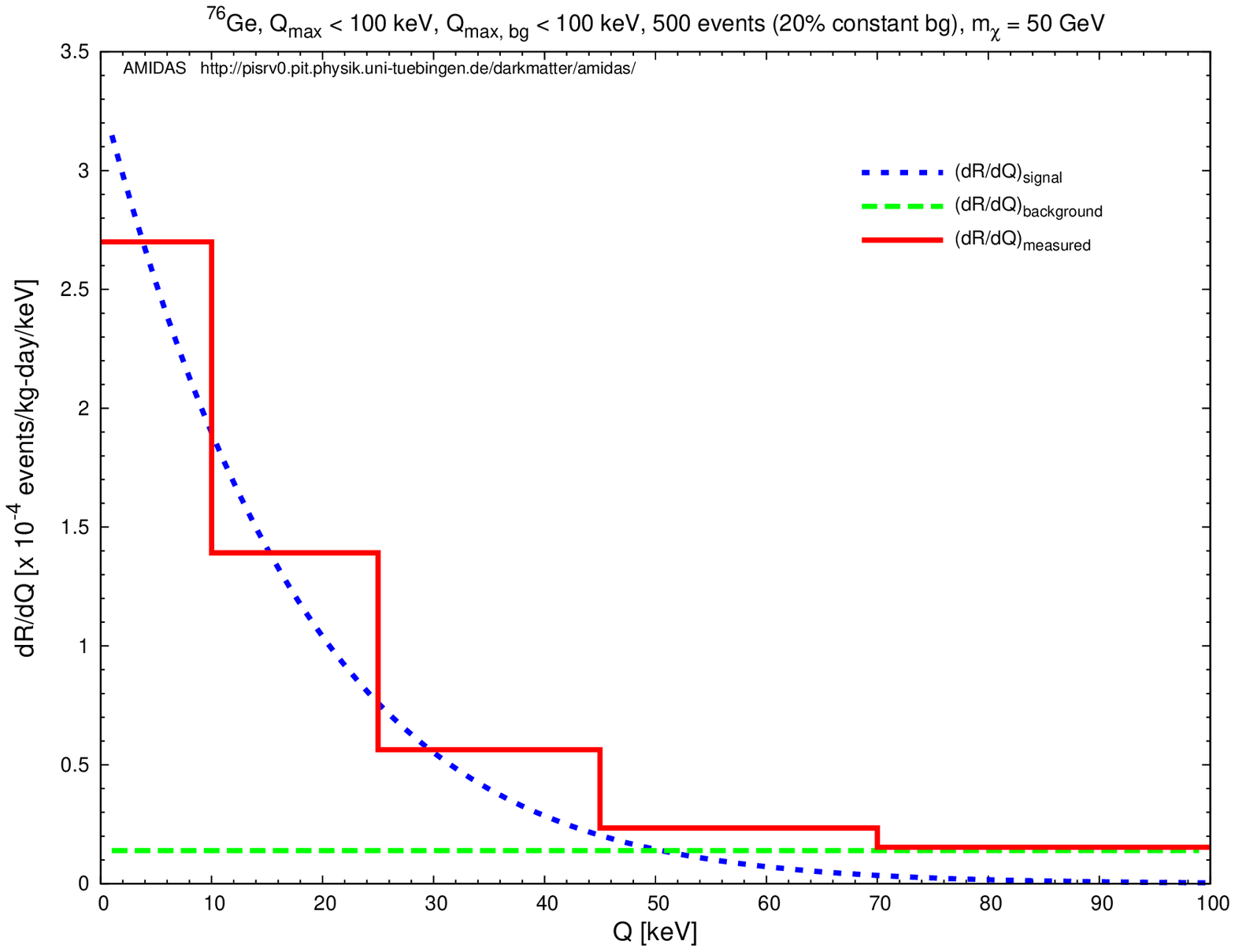} \hspace{-1.1cm}
 \includegraphics[width=9.8cm]{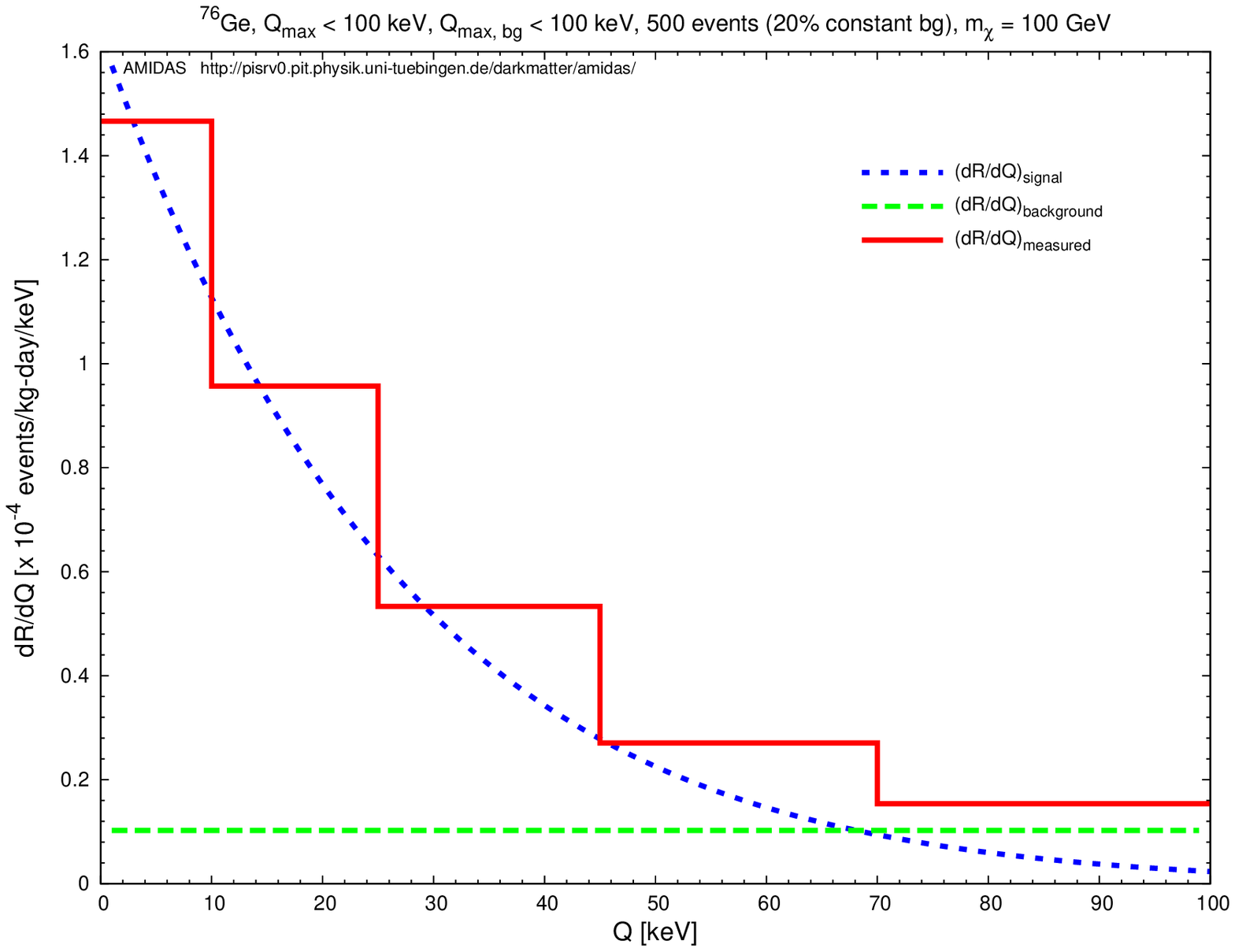} \hspace*{-1.6cm} \\
 \vspace{1cm}
 \hspace*{-1.6cm}
 \includegraphics[width=9.8cm]{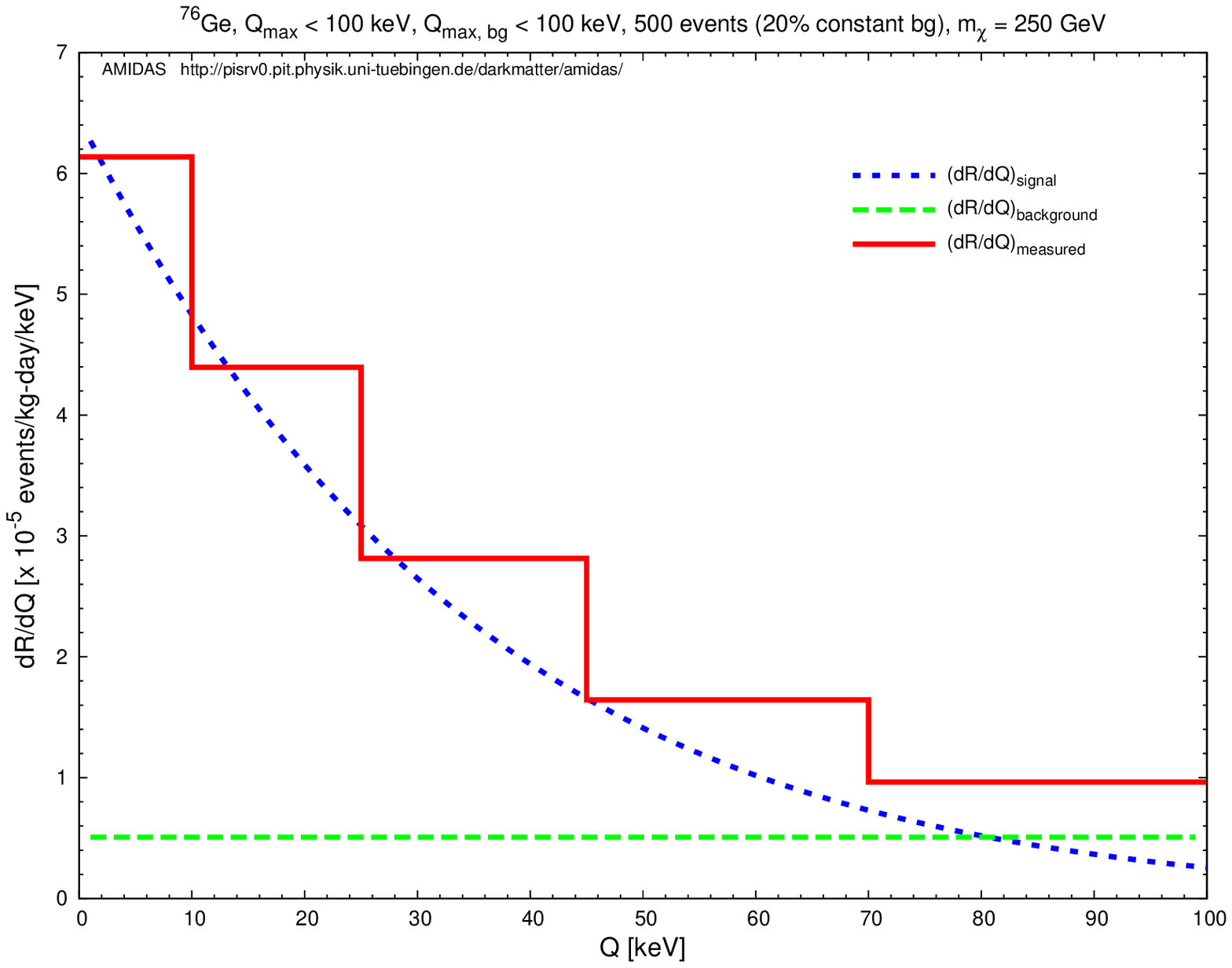} \hspace{-1.1cm}
 \includegraphics[width=9.8cm]{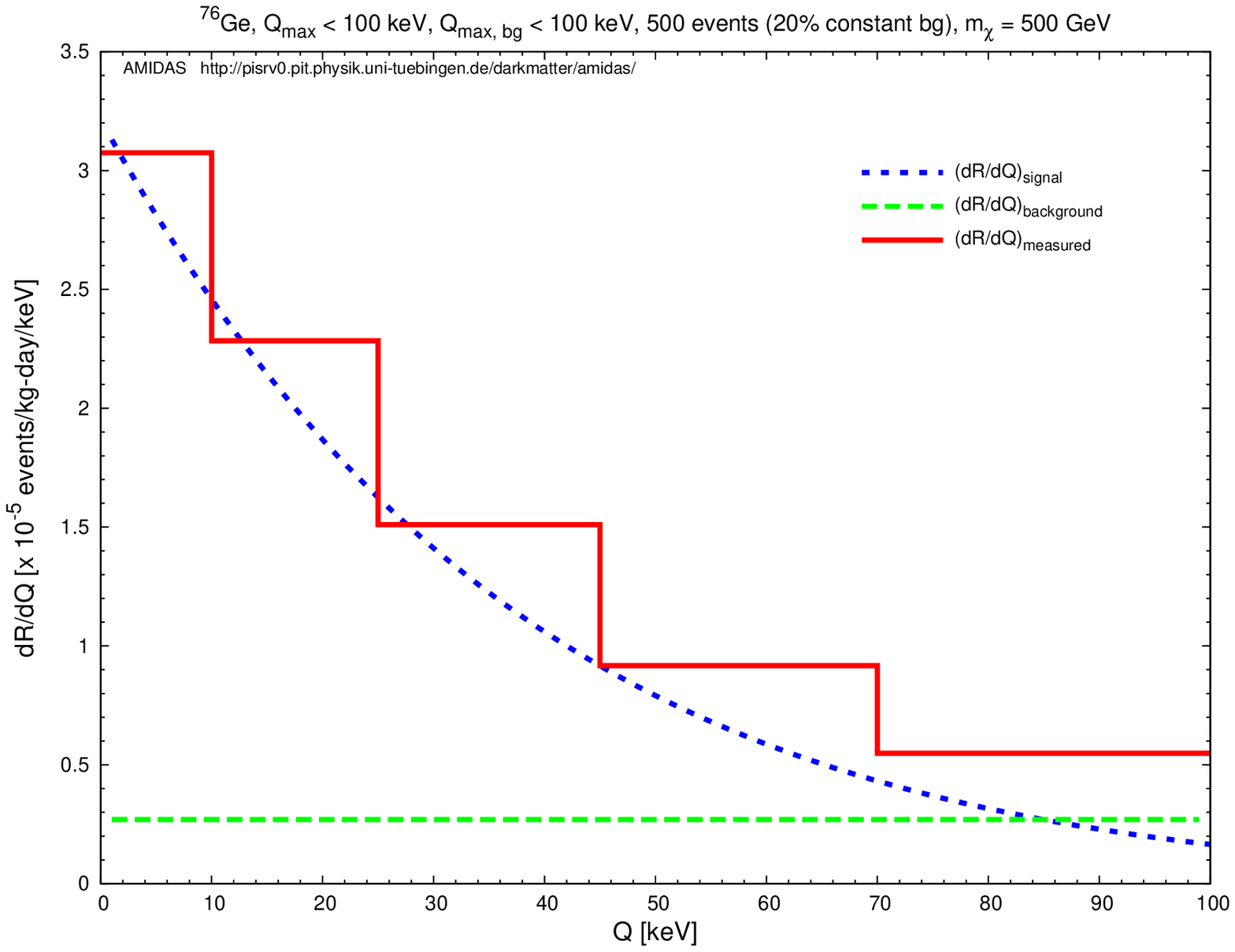} \hspace*{-1.6cm} \\
}
%\vspace{-0.25cm}
\end{center}
\caption{
 As in Figs.~\ref{fig:dRdQ-bg-ex-Ge-000-100-20},
 except that
 the constant background spectrum
 in Eq.~(\ref{eqn:dRdQ_bg_const})
 has been used.
 See the text for further details.
}
\label{fig:dRdQ-bg-const-Ge-000-100-20}
\end{figure}

 In Figs.~\ref{fig:dRdQ-bg-ex-Ge-000-100-20}
 we show measured energy spectra (solid red histograms)
 for a $\rmXA{Ge}{76}$ target
 with six different WIMP masses:
 10, 25, 50, 100, 250, and 500 GeV
 based on Monte Carlo simulations.
 The dotted blue curves are
 the elastic WIMP--nucleus scattering spectra
 for the shifted Maxwellian velocity distribution
 given in Eq.~(\ref{eqn:f1v_sh})
 with $v_0 = 220$ km/s, $\ve = 1.05 \~ v_0$,%
\footnote{
 The time dependence of the Earth's velocity
 in the Galactic frame,
 the second term of $\ve(t)$ in Eq.~(\ref{eqn:ve}),
 has been ignored.
}
 and $\vesc = 700$ km/s
 and the Woods--Saxon elastic nuclear form factor
 in Eq.~(\ref{eqn:FQ_SI_WS}).
 The dashed green curves are
 the exponential background spectra
 given in Eq.~(\ref{eqn:dRdQ_bg_ex}),
 which have been normalized so that
 the ratios of the areas under these background spectra
 to those under the (dotted blue) WIMP scattering spectra
 are equal to the background--signal ratio
 in the whole data sets
 (i.e., 20\% backgrounds to 80\% signals
  shown in Figs.~\ref{fig:dRdQ-bg-ex-Ge-000-100-20}).
 The experimental threshold energy
 has been assumed to be negligible
 and the maximal cut--off energy
 is set as 100 keV.
 5,000 experiments with 500 total events on average
 in each experiment have been simulated.

 The measured energy spectra (solid red histograms)
 shown in Figs.~\ref{fig:dRdQ-bg-ex-Ge-000-100-20}
 are averaged over the simulated experiments.
 Five bins with linearly increased bin widths
 have been used for binning
 generated signal and background events.
 As argued in Ref.~\cite{DMDDf1v},
 for reconstructing the one--dimensional
 WIMP velocity distribution function,
 this unusual, particular binning has been chosen
 in order to accumulate more events
 in high energy ranges
 and thus to reduce the statistical uncertainties
 in high velocity ranges.
 However,
 as we will show later,
 for the determination of the WIMP mass,
 one needs either events in the {\em first} energy bin
 or {\em all} events in the whole data set.
 Hence,
 there is in practice no difference
 between using an equal bin width for all bins
 or the (linearly) increased bin widths.

 Note here that,
 firstly,
 the possible energy ranges
 in which residue background events exist
 (the background windows)
 have been assumed to be
 the same as the entire experimental possible energy ranges
 (e.g.,~between 0 and 100 keV for simulations
  shown in Figs.~\ref{fig:dRdQ-bg-ex-Ge-000-100-20}).
 Secondly,
 the actual numbers of signal and background events
 in each simulated experiment
 are Poisson--distributed around their expectation values
 {\em independently}.
 This means that,
 for example,
 for simulations shown in Figs.~\ref{fig:dRdQ-bg-ex-Ge-000-100-20}
 we generate 400 (100) events on average for WIMP signals (backgrounds)
 and the total event number recorded in one experiment
 is then the sum of these two numbers.
 Thirdly,
 for the simulations demonstrated here
 as well as in the next section,
 we assumed that
 all experimental systematic uncertainties
 as well as the uncertainty on
 the measurement of the recoil energy
 could be ignored.
 The energy resolution of most existing detectors
 is so good that its error can be neglected
 compared to the statistical uncertainty
 for the foreseeable future
 with pretty few events.

 In Figs.~\ref{fig:dRdQ-bg-ex-Ge-000-100-20}
 it can be found that,
 as mentioned earlier,
 the shape of the WIMP scattering spectrum
 depends highly on the WIMP mass:
 for light WIMPs ($\mchi~\lsim~50$ GeV),
 the recoil spectra drop sharply with increasing recoil energies,
 while for heavy WIMPs ($\mchi~\gsim~100$ GeV),
 the spectra become flatter.
 In contrast,
 the exponential background spectra shown here
 depend only on the target mass
 and are rather {\em flatter} ({\em sharper})
 for {\em light} ({\em heavy}) WIMP masses
 compared to the WIMP scattering spectra.
 This means that,
 once input WIMPs are {\em light} ({\em heavy}),
 background events would contribute relatively more to
 {\em high} ({\em low}) energy ranges,
 and, consequently,
 the measured energy spectra
 would mimic scattering spectra
 induced by {\em heavier} ({\em lighter}) WIMPs.

 As a comparison,
 in Figs.~\ref{fig:dRdQ-bg-const-Ge-000-100-20}
 we generate background events
 with the constant spectrum
 given in Eq.~(\ref{eqn:dRdQ_bg_const}).
 It can be seen clearly that,
 since the background spectrum now
 is flatter for all WIMP masses,
 background events contribute always relatively more to
 {\em high} energy ranges,
 and, therefore,
 the measured energy spectra
 would always mimic scattering spectra
 induced by {\em heavier} WIMPs.
\section{Reconstruction of the WIMP mass}
 In this section
 we first review the model--independent method
 for determining the WIMP mass
 introduced in Refs.~\cite{DMDDmchi-SUSY07, DMDDmchi}.
 Then we demonstrate some numerical results
 of the reconstructed WIMP mass
 by using mixed data sets from WIMP signals and background events
 based on Monte Carlo simulations.
\subsection{Model--independent determination of the WIMP mass}
 Here we review briefly
 the model--independent data analysis procedure
 for the determination of the WIMP mass
 by using two experimental data sets
 with different target nuclei.
 Detailed derivations and discussions
 can be found in Refs.~\cite{DMDDmchi-SUSY07, DMDDmchi}.
\subsubsection{Basic expressions for determining the WIMP mass}
 In the earlier work \cite{DMDDf1v},
 it was found that
 the normalized one--dimensional velocity distribution function
 of incident WIMPs can be solved
 from Eq.~(\ref{eqn:dRdQ}) directly
 and, consequently,
 its generalized moments can be estimated by \cite{DMDDmchi}
\beqn
    \expv{v^n}(v(\Qmin), v(\Qmax))
 \= \int_{v(\Qmin)}^{v(\Qmax)} v^n f_1(v) \~ dv
    \non\\
 \= \alpha^n
    \bfrac{2 \Qmin^{(n+1)/2} r(\Qmin) / \FQmin + (n+1) I_n(\Qmin, \Qmax)}
          {2 \Qmin^{   1 /2} r(\Qmin) / \FQmin +       I_0(\Qmin, \Qmax)}
\~.
\label{eqn:moments}
\eeqn
 Here $v(Q) = \alpha \sqrt{Q}$,
 $Q_{\rm (min, max)}$ are the experimental
 minimal and maximal cut--off energies,
\beq
        r(\Qmin)
 \equiv \adRdQ_{{\rm expt},\~Q = \Qmin}
\label{eqn:rmin}
\eeq
 is an estimated value
 of the {\em measured} recoil spectrum
 $(dR / dQ)_{\rm expt}$ ({\em before}
 the normalization by the exposure $\cal E$) at $Q = \Qmin$,
%
%\footnote{
% Formulae needed for estimating
% $r(\Qmin)$, $I_n(\Qmin, \Qmax)$,
% and their statistical errors
% are given in the appendix.
%% can be found in Refs.~\cite{DMDDmchi, DMDDmchi-NJP}.
%}
%
 and $I_n(\Qmin, \Qmax)$ can be estimated through the sum:
\beq
   I_n(\Qmin, \Qmax)
 = \sum_a \frac{Q_a^{(n-1)/2}}{F^2(Q_a)}
\~,
\label{eqn:In_sum}
\eeq
 where the sum runs over all events in the data set
 that satisfy $Q_a \in [\Qmin, \Qmax]$.

 By requiring that
 the values of a given moment of $f_1(v)$
 estimated by Eq.~(\ref{eqn:moments})
 from two experiments
 with different target nuclei, $X$ and $Y$, agree,
 $\mchi$ appearing in the prefactor $\alpha^n$
 on the right--hand side of Eq.~(\ref{eqn:moments})
 has been solved as \cite{DMDDmchi-SUSY07}:
%
%\footnote{
% Formulae needed for estimating
% the statistical errors on
% $\left. \mchi \right|_{\Expv{v^n}}$ and
% $\left. \mchi \right|_\sigma$
% are given in the appendix.
%% can be found in Refs.~\cite{DMDDmchi, DMDDmchi-NJP}.
%}:
%
\beq
   \left. \mchi \right|_{\Expv{v^n}}
 = \frac{\sqrt{\mX \mY} - \mX (\calR_{n, X} / \calR_{n, Y})}
        {\calR_{n, X} / \calR_{n, Y} - \sqrt{\mX / \mY}}
\~,
\label{eqn:mchi_Rn}
\eeq
 where
\beqn
        \calR_{n, X}
 \equiv \bfrac{2 \QminX^{(n+1)/2} r_X(\QminX) / \FQminX + (n+1) \InX}
              {2 \QminX^{   1 /2} r_X(\QminX) / \FQminX +       \IzX}^{1/n}
\~,
\label{eqn:RnX_min}
\eeqn
 and $\calR_{n, Y}$ can be defined analogously.
 Here $n \ne 0$,
 $m_{(X, Y)}$ and $F_{(X, Y)}(Q)$
 are the masses and the form factors of the nucleus $X$ and $Y$,
 respectively,
 and $r_{(X, Y)}(Q_{{\rm min}, (X, Y)})$
 refer to the counting rates for detectors $X$ and $Y$
 at the respective lowest recoil energies included in the analysis.
 Note that,
 firstly,
 the general expression (\ref{eqn:mchi_Rn}) can be used
 either for spin--independent or for spin--dependent scattering,
 one only needs to choose different form factors
 under different assumptions.
 Secondly,
 the form factors in the estimate of $\InX$ and $\InY$
 using Eq.~(\ref{eqn:In_sum}) are also different.

 On the other hand,
 by using the theoretical prediction that
 the SI WIMP--nucleus cross section
 dominates,
 and the fact that
 the integral over the one--dimensional WIMP velocity distribution
 on the right--hand side of Eq.~(\ref{eqn:dRdQ})
 is the minus--first moment of this distribution,
 which can be estimated by Eq.~(\ref{eqn:moments}) with $n = -1$,
 one can easily find that \cite{DMDDmchi}
\beq
   \rho_0 |f_{\rm p}|^2
 = \frac{\pi}{4 \sqrt{2}} \afrac{\mchi + \mN}{\calE A^2 \sqrt{\mN}}
   \bbrac{\frac{2 \Qmin^{1/2} r(\Qmin)}{\FQmin} + I_0}
\~.
\label{eqn:rho0_fp2}
\eeq
 Note that
 the exposure of the experiment, $\calE$,
 appears in the denominator.
 Since the unknown factor $\rho_0 |f_{\rm p}|^2$
 on the left--hand side above
 is identical for different targets,
 it leads to a second expression for determining $\mchi$
 \cite{DMDDmchi}:
\beq
   \left. \mchi \right|_\sigma
 = \frac{\abrac{\mX / \mY}^{5/2} \mY - \mX (\calR_{\sigma, X} / \calR_{\sigma, Y})}
        {\calR_{\sigma, X} / \calR_{\sigma, Y} - \abrac{\mX / \mY}^{5/2}}
\~.
\label{eqn:mchi_Rsigma}
\eeq
 Here $m_{(X, Y)} \propto A_{(X, Y)}$ has been assumed,
\beq
        \calR_{\sigma, X}
 \equiv \frac{1}{\calE_X}
        \bbrac{\frac{2 \QminX^{1/2} r_X(\QminX)}{\FQminX} + \IzX}
\~,
\label{eqn:RsigmaX_min}
\eeq
 and similarly for $\calR_{\sigma, Y}$.
\subsubsection{\boldmath$\chi^2$--fitting}
 In order to yield the best--fit WIMP mass
 as well as to minimize its statistical uncertainty
 by combining the estimators for different $n$
 in Eq.~(\ref{eqn:mchi_Rn}) with each other
 and with the estimator in Eq.~(\ref{eqn:mchi_Rsigma}),
 a $\chi^2$ function has been introduced as
 \cite{DMDDmchi}
\beq
   \chi^2(\mchi)
 = \sum_{i, j}
   \abrac{f_{i, X} - f_{i, Y}} {\cal C}^{-1}_{ij} \abrac{f_{j, X} - f_{j, Y}}
\~,
\label{eqn:chi2}
\eeq
 where
\cheqna
\beqn
           f_{i, X}
 \eqnequiv \alpha_X^i
           \bfrac{  2 Q_{{\rm min}, X}^{(i+1)/2} r_X(\Qmin) / F^2_X(Q_{{\rm min}, X})
                  + (i+1) I_{i, X}}
                 {  2 Q_{{\rm min}, X}^{   1 /2} r_X(\Qmin) / F^2_X(Q_{{\rm min}, X})
                  +       \IzX}
           \afrac{1}{300~{\rm km/s}}^i
%           \non\\
%           \non\\
% \=        \afrac{\alpha_X {\cal R}_{i, X}}{300~{\rm km/s}}^{i}
\~,
\label{eqn:fiXa}
\eeqn
 for $i = -1,~1,~2,~\dots,~n_{\rm max}$, and
\cheqnb
\beqn
           f_{n_{\rm max}+1, X}
 \eqnequiv \calE_X
           \bfrac{A_X^2}
                 {  2 Q_{{\rm min}, X}^{1/2} r_X(\Qmin) / F^2_X(Q_{{\rm min}, X})
                  + \IzX}
           \afrac{\sqrt{\mX}}{\mchi + \mX}
%           \non\\
%           \non\\
% \=        \frac{A_X^2}{\calR_{\sigma, X}} \afrac{\sqrt{\mX}}{\mchi + \mX}
\~;
\label{eqn:fiXb}
\eeqn
\cheqn
 the other $n_{\rm max} + 2$ functions $f_{i, Y}$
 can be defined analogously.
 Here $n_{\rm max}$ determines the highest moment of $f_1(v)$
 that is included in the fit.
 The $f_i$ are normalized such that
 they are dimensionless and very roughly of order unity
 in order to alleviate numerical problems
 associated with the inversion of their covariance matrix.
 Note that
 the first $n_{\rm max} + 1$ fit functions
 depend on $\mchi$ only through the overall factor $\alpha$
 and that
 $\mchi$ in Eqs.~(\ref{eqn:fiXa}) and (\ref{eqn:fiXb})
 is now a fit parameter,
 which may differ from the true value of the WIMP mass.
 Finally,
 $\cal C$ in Eq.~(\ref{eqn:chi2}) is the total covariance matrix.
 Since the $X$ and $Y$ quantities
 are statistically completely independent,
 $\cal C$ can be written as a sum of two terms:
%
%\footnote{
% Formulae needed for estimating the entries of $\cal C$
% are given in the appendix.
%% can be found in Refs.~\cite{DMDDmchi, DMDDmchi-NJP}.
%}:
%
\beq
   {\cal C}_{ij}
 = {\rm cov}\abrac{f_{i, X}, f_{j, X}} + {\rm cov}\abrac{f_{i, Y}, f_{j, Y}}
\~.
\label{eqn:Cij}
\eeq
\subsubsection{Matching the cut--off energies}
 The basic requirement of the expressions for determining $\mchi$
 given in Eqs.~(\ref{eqn:mchi_Rn}) and (\ref{eqn:mchi_Rsigma}) is that,
 from two experiments with different target nuclei,
 the values of a given moment of the WIMP velocity distribution
 estimated by Eq.~(\ref{eqn:moments}) should agree.
 This means that
 the upper cuts on $f_1(v)$ in two data sets
 should be (approximately) equal%
\footnote{
 Here the threshold energies
 have been assumed to be negligibly small.
}.
 Since $v_{\rm cut} = \alpha \sqrt{Q_{\rm max}}$,
 it requires that \cite{DMDDmchi}
\beq
   Q_{{\rm max}, Y}
 = \afrac{\alpha_X}{\alpha_Y}^2 Q_{{\rm max}, X}
\~.
\label{eqn:match}  
\eeq
 Note that
 $\alpha$ defined in Eq.~(\ref{eqn:alpha})
 is a function of the true WIMP mass.
 Thus this relation for matching optimal cut--off energies
 can be used only if $\mchi$ is already known.
 One possibility to overcome this problem is
 to fix the cut--off energy of the experiment with the heavier target,
 minimize the $\chi^2(\mchi)$ function
 defined in Eq.~(\ref{eqn:chi2}),
 and then estimate the cut--off energy for the lighter nucleus
 by Eq.~(\ref{eqn:match}) algorithmically \cite{DMDDmchi}.
\subsection{Reconstructing \boldmath$\mchi$
            by using data sets with background events}
 In this subsection
 we show some numerical results
 of the reconstruction of the WIMP mass
 with mixed data sets
 from WIMP--induced and background events
 by means of the model--independent method
 described in the previous subsection.
 The upper and lower bounds on the reconstructed WIMP mass
 are estimated from the requirement that
 $\chi^2$ exceeds its minimum by 1.%
\footnote{
 Note that,
 rather than the mean values,
 the (bounds on the) reconstructed WIMP mass
 are always the {\em median} values
 of the simulated results.
}
 As in Ref.~\cite{DMDDmchi},
 $\rmXA{Si}{28}$ and $\rmXA{Ge}{76}$
 have been chosen as two target nuclei.
 The scattering cross section $\sigma_0$ in Eq.~(\ref{eqn:calA})
 has been assumed to be dominated by
 the spin--independent WIMP--nucleus interaction.
 The experimental threshold energies of two experiments
 have been assumed to be negligible
 and the maximal cut--off energies
 are set the same as 100 keV.
 2 $\times$ 5,000 experiments have been simulated.
 In order to avoid large contributions
 from very few events in high energy ranges
 to the higher moments \cite{DMDDf1v},
 only the moments up to $n_{\rm max} = 2$
 were included in the $\chi^2$ fit.
\subsubsection{With the exponential background spectrum}
\begin{figure}[t!]
\begin{center}
\imageswitch{
\begin{picture}(15,10.5)
\put(0,0){\framebox(15,10.5) {mchi-SiGe-ex-000-100-050}}
\end{picture}}
{\includegraphics[width=15cm]{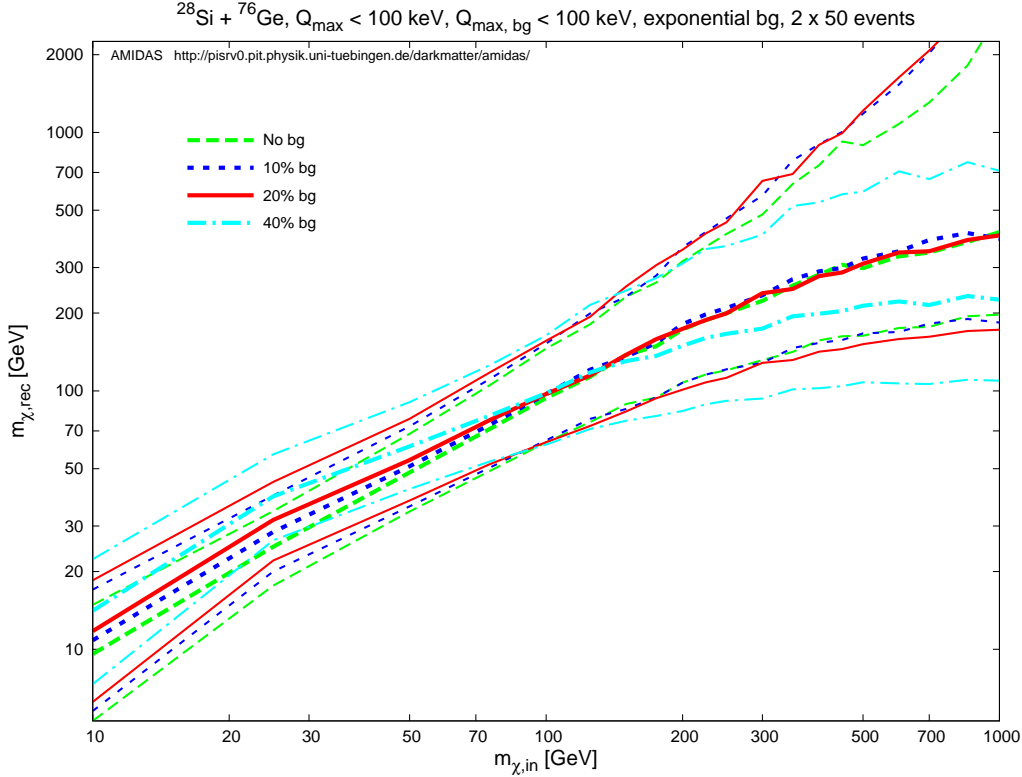} \\
}
\vspace{-0.5cm}
\end{center}
\caption{
 The reconstructed WIMP mass
 and the lower and upper bounds of
 the 1$\sigma$ statistical uncertainty
 with mixed data sets
 from WIMP--induced and background events
 as functions of the input WIMP mass.
 $\rmXA{Si}{28}$ and $\rmXA{Ge}{76}$
 have been chosen as two target nuclei.
 The background ratios shown here
 are no background (dashed green curves),
%  5\% (dotted magenta curves),
 10\% (long--dotted blue curves),
 20\% (solid red curves),
 and 40\% (dash--dotted cyan curves)
 background events in the whole data sets
 in the experimental energy ranges
 between 0 and 100 keV.
 Each experiment contains 50 {\em total} events
 on average before cuts on $\Qmax$
 for the experiments with the Si target;
 {\em all} of these events are treated as WIMP signals.
 Other parameters are as
 in Figs.~\ref{fig:dRdQ-bg-ex-Ge-000-100-20}.
 See the text for further details.
}
\label{fig:mchi-SiGe-ex-000-100-050}
\end{figure}

 Fig.~\ref{fig:mchi-SiGe-ex-000-100-050}
 shows the reconstructed WIMP mass
 and the lower and upper bounds of
 the 1$\sigma$ statistical uncertainty
 with mixed data sets
 from WIMP--induced and background events
 as functions of the input WIMP mass.
 As in Figs.~\ref{fig:dRdQ-bg-ex-Ge-000-100-20},
 the exponential background spectrum has been used
 and the background windows are set as
 the same as the experimental possible energy ranges,
 i.e., between 0 and 100 keV
 for both experiments.
 The background ratios shown here
 are no background (dashed green curves),
%  5\% (dotted magenta curves),
 10\% (long--dotted blue curves),
 20\% (solid red curves),
 and 40\% (dash--dotted cyan curves)
 background events in the whole data sets.
 Each experiment contains 50 {\em total} events
 on average before cuts on $\Qmax$
 for the experiments with the Si target.
 Remind that
 {\em all} events recorded in our data sets
 are treated as WIMP signals in the analysis,
 although statistically we know that
 a fraction of these events could be backgrounds.

 It can be seen clearly that,
 for {\em light} WIMP masses ($\mchi~\lsim~100$ GeV),
 the larger the fraction of background events in the data sets,
 the {\em heavier} the reconstructed WIMP masses
 as well as the statistical uncertainty intervals.
 This is caused directly by
 the background contribution to {\em high} energy ranges
 shown in Figs.~\ref{fig:dRdQ-bg-ex-Ge-000-100-20}.
 As discussed in Sec.~2.3,
 the background spectrum is relatively {\em flatter}
 compared to the scattering spectrum
 induced by {\em light} WIMPs,
 and the energy spectrum of all recorded events
 would thus mimic a scattering spectrum induced
 by WIMPs with a {\em relatively heavier} mass.
 Not surprisingly,
 the larger the background ratio,
 the more the background contribution
 to high energy ranges,
 and, consequently,
 the more strongly {\em overestimated}
 the reconstructed WIMP masses
 as well as the statistical uncertainty intervals.

 In contrast,
 for {\em heavy} WIMP masses ($\mchi~\gsim~100$ GeV),
 Fig.~\ref{fig:mchi-SiGe-ex-000-100-050}
 does not show very clearly but a tendency%
\footnote{
 Since
 for heavy input WIMP masses
 the reconstructed values
 are {\em systematically underestimated},
 probably due to the statistical fluctuation
 with pretty few ($\sim 50$) events
 discussed later.
}
 that
 the larger the fraction of background events,
 the {\em lighter} the reconstructed WIMP masses
 as well as the statistical uncertainty intervals.
 This is now caused by
 the background contribution to {\em low} energy ranges
 shown in Figs.~\ref{fig:dRdQ-bg-ex-Ge-000-100-20}.
 As discussed in the previous section,
 the background spectrum is relatively {\em sharper}
 compared to the scattering spectrum
 induced by {\em heavy} WIMPs,
 and the energy spectrum of all recorded events
 would thus mimic a scattering spectrum induced
 by WIMPs with a {\em relatively lighter} mass.
 Moreover,
 the larger the background ratio,
 the more the background contribution
 to low energy ranges,
 and, consequently,
 the more strongly {\em underestimated}
 the reconstructed WIMP masses
 as well as the statistical uncertainty intervals.

\begin{figure}[p!]
\begin{center}
\imageswitch{
\begin{picture}(15,22)
\put(0,11.5){\framebox(15,10.5){mchi-SiGe-ex-000-050-050}}
\put(0, 0  ){\framebox(15,10.5){mchi-SiGe-ex-050-100-050}}
\end{picture}}
{\includegraphics[width=15cm]{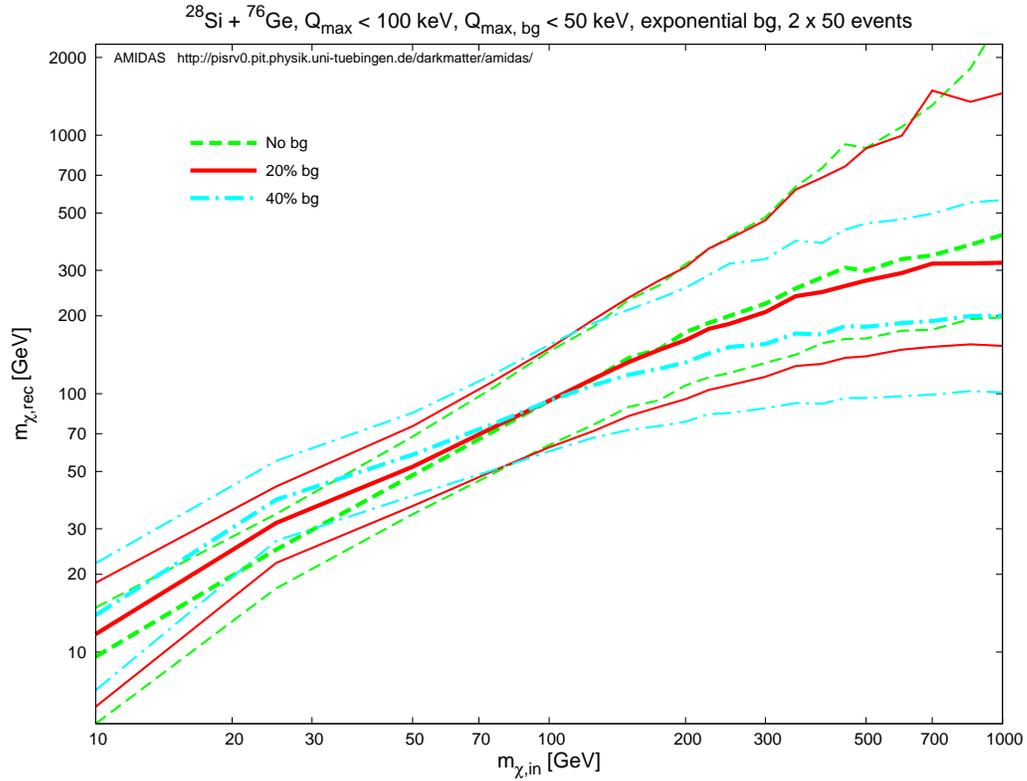}  \\ \vspace{0.75cm}
 \includegraphics[width=15cm]{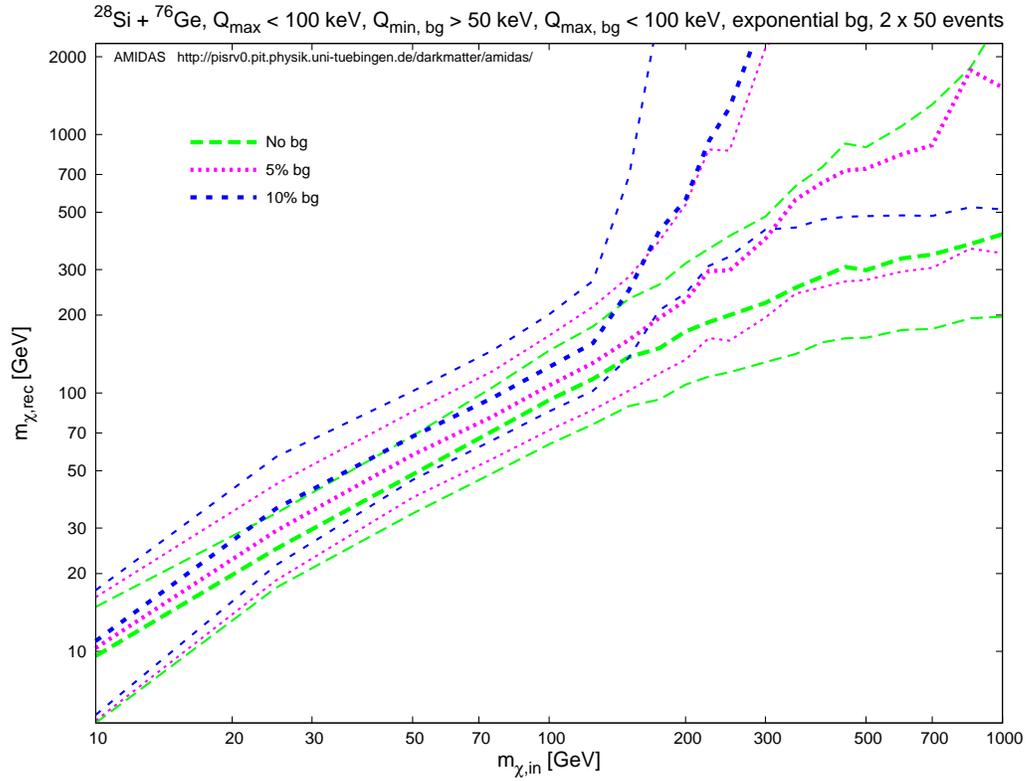}  \\
}
\vspace{-0.25cm}
\end{center}
\caption{
 As in Fig.~\ref{fig:mchi-SiGe-ex-000-100-050},
 except that
 the background window in each experiment
 have been set as
  0 --  50 keV (upper) and
 50 -- 100 keV (lower).
 Note that
 the background ratios shown here
 are % no background (dashed green curves),
 20\% (solid red curves)
 and 40\% (dash--dotted cyan curves) in the upper frame,
 whereas 5\% (dotted magenta curves)
 and 10\% (long--dotted blue curves) in the lower frame.
}
\label{fig:mchi-SiGe-ex-000-050-100-050}
\end{figure}

 Nevertheless,
 from Fig.~\ref{fig:mchi-SiGe-ex-000-100-050}
 it can be found that,
 with $\sim$ 20\% residue background events
 in the analyzed data sets,
 the true values of the WIMP mass
 can still fall in the middle of
 the 1$\sigma$ statistical uncertainty band
 and one could thus in principle
 reconstruct the WIMP mass pretty well;
 if WIMPs are light ($\mchi~\lsim~200$ GeV),
 the maximal acceptable fraction of
 residue background events
 could even be as large as $\sim$ 40\%.
 For a WIMP mass of 100 GeV with
 20\% background events in the data sets,
 the reconstructed WIMP mass and
 the statistical uncertainty
 are $\sim 97~{\rm GeV}\~^{+61\%}_{-35\%}$,
 compared to $\sim 94~{\rm GeV}\~^{+55\%}_{-33\%}$
 for background--free data sets;
 for a lighter WIMP mass of 50 GeV,
 the reconstructed WIMP mass and
 the statistical uncertainty
 change from $\sim 48~{\rm GeV}\~^{+41\%}_{-29\%}$ (background--free),
 to $\sim 54~{\rm GeV}\~^{+44\%}_{-30\%}$ (20\% background),
 and $\sim 61~{\rm GeV}\~^{+48\%}_{-32\%}$ (40\% background).

 On the other hand,
 considering different efficiencies of
 discrimination ability
 against different background sources
 in different energy ranges
 in different experiments,
 in Figs.~\ref{fig:mchi-SiGe-ex-000-050-100-050}
 we shrink the background window in each experiment to
 a relatively lower range
 between 0 and 50 keV (upper) and
 a relatively higher range
 between 50 and 100 keV (lower)%
\footnote{
 Note that
 here we do {\em not} mean that
 in other energy ranges background events do not exist;
 in contrast,
 we want to study what could happen
 once our background discrimination,
 caused by some natural or even artificial reasons,
 are {\em worse} in these energy ranges than others and
 {\em more} background events could thus survive.%
}.
 Since our background spectrum is exponential,
 for the case shown in Fig.~\ref{fig:mchi-SiGe-ex-000-100-050},
 only very few background events
 could be observed in the energy range
 between 50 and 100 keV.
 Hence,
 for the case with the background window
 only in the {\em low} energy range,
 not surprisingly,
 the results of the reconstructed WIMP mass
 shown in the upper frame of
 Figs.~\ref{fig:mchi-SiGe-ex-000-050-100-050}
 should not differ very much from those
 shown in Fig.~\ref{fig:mchi-SiGe-ex-000-100-050}.
 However,
 due to the little bit more contribution
 to the {\em low} energy range from background events,
 {\em all} the reconstructed WIMP masses shown here
 are somehow {\em lighter} than those
 shown in Fig.~\ref{fig:mchi-SiGe-ex-000-100-050}.
 Hence,
 with $\sim$ 20\% residue background events
 in low experimental possible energy ranges,
 one could in principle reconstruct the WIMP mass
 with a 1$\sigma$ statistical uncertainty as
 $\sim 94~{\rm GeV}\~^{+59\%}_{-34\%}$
 (for a WIMP mass of 100 GeV) or
 $\sim 52~{\rm GeV}\~^{+44\%}_{-30\%}$
 (for a WIMP mass of 50 GeV).

 In contrast,
 since the WIMP scattering spectrum is
 in principle approximately exponential
 and thus only (very) few WIMP--induced events
 could be observed in high energy ranges,
 if we have background windows
 in only high experimental possible energy ranges,
 the (pretty large) contributions from background events
 could cause (strong) {\em overestimates}
 of the reconstructed WIMP masses.
 It is even worse for large WIMP masses
 ($\mchi~\gsim~100$ GeV)%
\footnote{
 Note that
 the plateau of the lower bound of
 the statistical uncertainty
 in the case of a 10\% background ratio
 for heavy WIMP masses
 ($\mchi~\gsim~300$ GeV) 
 should be caused by our setup
 for the upper cut--off of the reconstructed WIMP mass
 of 3000 GeV in the simulations. 
}.
 Nevertheless,
 as shown in the lower frame of
 Figs.~\ref{fig:mchi-SiGe-ex-000-050-100-050},
 with $\sim$ 5\% residue background events
 observed only in high energy ranges,
 one could in principle still estimate the WIMP mass
 with a 1$\sigma$ statistical uncertainty as
 $\sim 107~{\rm GeV}\~^{+56\%}_{-33\%}$
 (for an input WIMP mass of 100 GeV) or
 $\sim 58~{\rm GeV}\~^{+47\%}_{-32\%}$
 (for an input WIMP mass of 50 GeV).

 Our results shown in
 Figs.~\ref{fig:mchi-SiGe-ex-000-050-100-050}
 indicate that
 a small fraction of background events
 in {\em low} energy ranges
 might not affect the reconstructed WIMP mass significantly.
 However,
 the WIMP mass could be (strongly) {\em overestimated}
 once the same (or even smaller) amount of background events
 exists in {\em high} energy ranges.
 In practice
 one simple way to reduce the overestimate
 induced by an excess of background events
 in high energy ranges might be
 checking the shape of measured recoil spectrum.
 However,
 considering some suggested modifications of
 the standard shifted Maxwellian velocity distribution,
 e.g., contributions from discrete ``streams''
 with (nearly) fixed velocities
 \cite{Sikivie, %Sikivie92, Sikivie03,
       Freese04,
       Natarajan} % Natarajan06, Natarajan07}
 or the ``late infall'' component in the velocity distribution
 with a velocity $v \sim \vesc$
 \cite{Sikivie, %Sikivie92, Sikivie03,
       Ling04,
       Natarajan}, % Natarajan06, Natarajan07},
 it should at least be very careful
 to reject any recoil event
 observed in high energy ranges artificially.

\begin{figure}[t!]
\begin{center}
\imageswitch{
\begin{picture}(15,10.5)
\put(0,0){\framebox(15,10.5) {mchi-SiGe-ex-000-100-500}}
\end{picture}}
{\includegraphics[width=15cm]{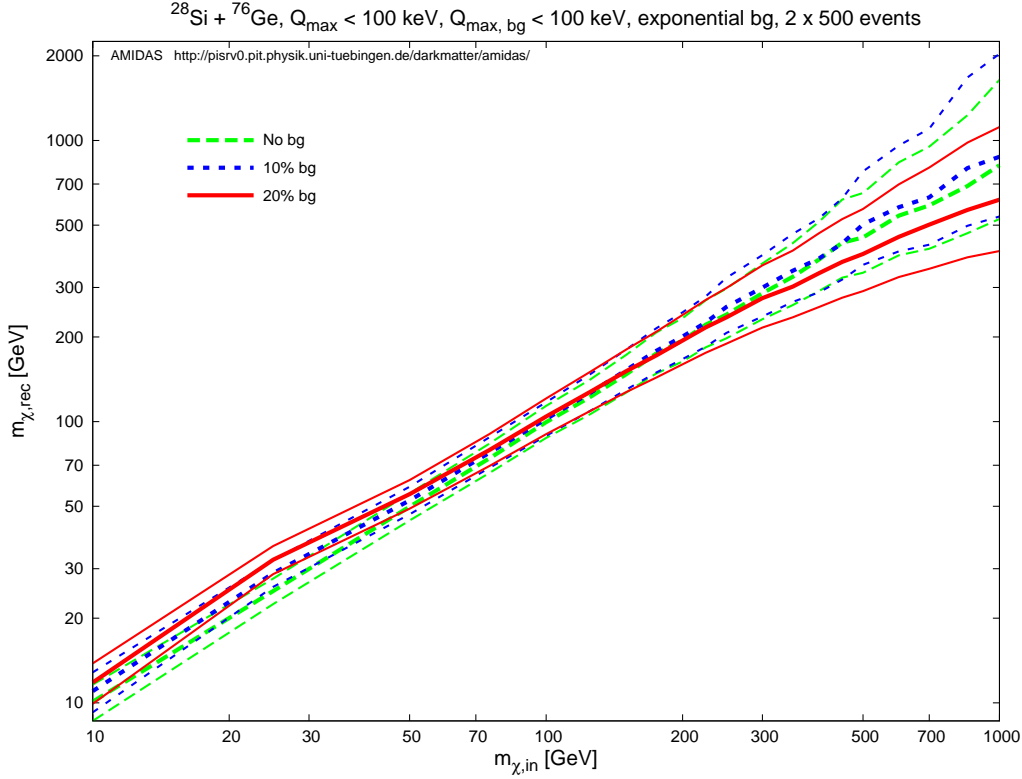} \\
}
\vspace{-0.5cm}
\end{center}
\caption{
 As in Fig.~\ref{fig:mchi-SiGe-ex-000-100-050},
 except that
 the expected number of total events
 in each experiment
 has been set as 500.
}
\label{fig:mchi-SiGe-ex-000-100-500}
\end{figure}
 In Fig.~\ref{fig:mchi-SiGe-ex-000-100-500}
 we rise the expected number of
 total events in each experiment
 by a factor of 10,
 to 500 events on average before cuts
 for the case that
 residue background events exist
 in the entire experimental possible energy ranges.
 As shown here,
 all statistical uncertainties shrink
 by a factor $\gsim~3$ compared to the results
 shown in Fig.~\ref{fig:mchi-SiGe-ex-000-100-050}.
 In addition,
 the underestimate of the reconstructed values
 of heavy input WIMP masses
 caused perhaps by the use of pretty few ($\sim 50$) events
 has been reduced with larger data sets;
 and,
 the tendency of the underestimate of
 the reconstructed WIMP mass
 for heavy WIMP masses ($\mchi~\gsim~100$ GeV)
 becomes more clearly.
 Finally,
 Fig.~\ref{fig:mchi-SiGe-ex-000-100-500}
 shows that,
 for the determination of the WIMP mass
 by using data sets of $\cal O$(500) total events,
 the maximal acceptable background ratio
 could be $\sim$ 10\%
 (i.e., $\cal O$(50) background events)
 or even $\sim$ 20\%,
 if WIMPs have a mass of $\cal O$(100 GeV).
\subsubsection{Statistical fluctuation}
 As discussed in Ref.~\cite{DMDDmchi},
 the statistical fluctuation of the reconstructed WIMP mass
 by the algorithmic procedure in the simulated experiments
 seems to be pretty problematic,
 in particular for heavier input WIMP masses.
 Moreover,
 as mentioned in the previous subsection,
 with only $\sim$ 50 total events in each experiment,
 the tendency of the underestimate of
 the reconstructed WIMP mass
 for heavier WIMP masses ($\mchi~\gsim~100$ GeV)
 seems not to be very clear.
 Hence,
 as done in Ref.~\cite{DMDDmchi},
 in order to study the statistical fluctuation of
 the reconstructed WIMP mass
 with different background ratios in our data sets,
 we consider in this subsection
 the estimator $\delta m$
 introduced in Ref.~\cite{DMDDmchi}:
\beq
\renewcommand{\arraystretch}{0.75}
   \delta m
 = \left\{
    \begin{array}{l c l}
     \D   1
        + \frac{m_{\chi, {\rm lo1}} - m_{\chi, {\rm in }}}
               {m_{\chi, {\rm lo1}} - m_{\chi, {\rm lo2}}}\~, & ~~~~~~ & %6
        {\rm if}~m_{\chi, {\rm in }} \leq m_{\chi, {\rm lo1}}\~;
     \\ & & \\
    \D   \frac {m_{\chi, {\rm rec}} - m_{\chi, {\rm in }}}
               {m_{\chi, {\rm rec}} - m_{\chi, {\rm lo1}}}\~, &        &
        {\rm if}~m_{\chi,{\rm lo1}} < m_{\chi, {\rm in }} < m_{\chi, {\rm rec}}\~;
    \\ & & \\
    \D   \frac {m_{\chi, {\rm rec}} - m_{\chi, {\rm in }}}
               {m_{\chi, {\rm hi1}} - m_{\chi, {\rm rec}}}\~, &        &
        {\rm if}~m_{\chi,{\rm rec}} < m_{\chi, {\rm in }} < m_{\chi, {\rm hi1}}\~;
    \\ & & \\
    \D   \frac {m_{\chi, {\rm hi1}} - m_{\chi, {\rm in }}}
               {m_{\chi, {\rm hi2}} - m_{\chi, {\rm hi1}}} - 1 \~, &   &
        {\rm if}~m_{\chi, {\rm in }} \geq m_{\chi,{\rm hi1}}\~.
    \end{array}
   \right.
\label{eqn:deltam}
\eeq
 Here $m_{\chi, {\rm in }}$ is the true (input) WIMP mass,
 $m_{\chi, {\rm rec}}$ its reconstructed value,
 $m_{\chi, {\rm lo1(2)}}$ are the $1 \~ (2) \~ \sigma$ lower bounds
 satisfying $\chi^2(m_{\chi, {\rm lo(1, 2)}}) = \chi^2(m_{\chi, {\rm rec}}) + 1 \~ (4)$,
 and $m_{\chi, {\rm hi1(2)}}$ are
 the corresponding $1 \~ (2) \~ \sigma$ upper bounds.

 The estimator $\delta m$ defined above
 indicates basically the {\em strength of the deviation}
 of the reconstructed WIMP mass
 from the true (input) value.
 If the reconstructed 1$\sigma$ lower and upper bounds
 on the WIMP mass in one simulated experiment
 cover the true value:
 $m_{\chi, {\rm lo1}} \le m_{\chi, {\rm in}} \le m_{\chi, {\rm hi1}}$,
 $\delta m$ is determined as
 the deviation of the ``reconstructed WIMP mass''
 from the true one
 in units of the difference
 between the reconstructed value
 and the 1$\sigma$ lower (upper) bound,
 once the reconstructed value
 is overestimated % ($m_{\chi, {\rm rec}} \ge m_{\chi, {\rm in}}$)
 (underestimated). % ($m_{\chi, {\rm rec}} \le m_{\chi, {\rm in}}$).
 However,
 if the true WIMP mass lies {\em outside} of
 the experimental 1$\sigma$ bounds
 (the reconstructed value is more strongly over-/underestimated),
 $\delta m$ is determined as
 the deviation of the ``1$\sigma$ lower (upper) bound''
 from the true WIMP mass
 in units of the difference
 between the 1$\sigma$ and 2$\sigma$ lower (upper) bounds.
 Note that,
 it has been found in Ref.~\cite{DMDDmchi}
 as well as in the results presented
 in the previous subsection that
 the uncertainty intervals of
 the median reconstructed WIMP mass
 are quite asymmetric;
 similarly,
 the distance between the 1$\sigma$ and 2$\sigma$ bounds
 can be quite different from
 the distance between the reconstructed value
 and the 1$\sigma$ bound \cite{DMDDmchi}.
 The definition of $\delta m$ in Eq.(\ref{eqn:deltam})
 takes these differences into account,
 and also keeps track of the sign of the deviation:
 if the reconstructed WIMP mass is overestimated (underestimated),
 $\delta m$ is positive (negative).
 Moreover,
 $|\delta m| \leq 1 \~ (2)$ if and only if
 the true WIMP mass lies between
 the experimental $1 \~ (2) \~ \sigma$ bounds.

\begin{figure}[p!]
\begin{center}
\imageswitch{
\begin{picture}(15,22)
\put(0,11.5){\framebox(15,10.5){mchi-dev-SiGe-ex-000-100-050}}
\put(0, 0  ){\framebox(15,10.5){mchi-dev-SiGe-ex-000-100-500}}
\end{picture}}
{\includegraphics[width=15cm]{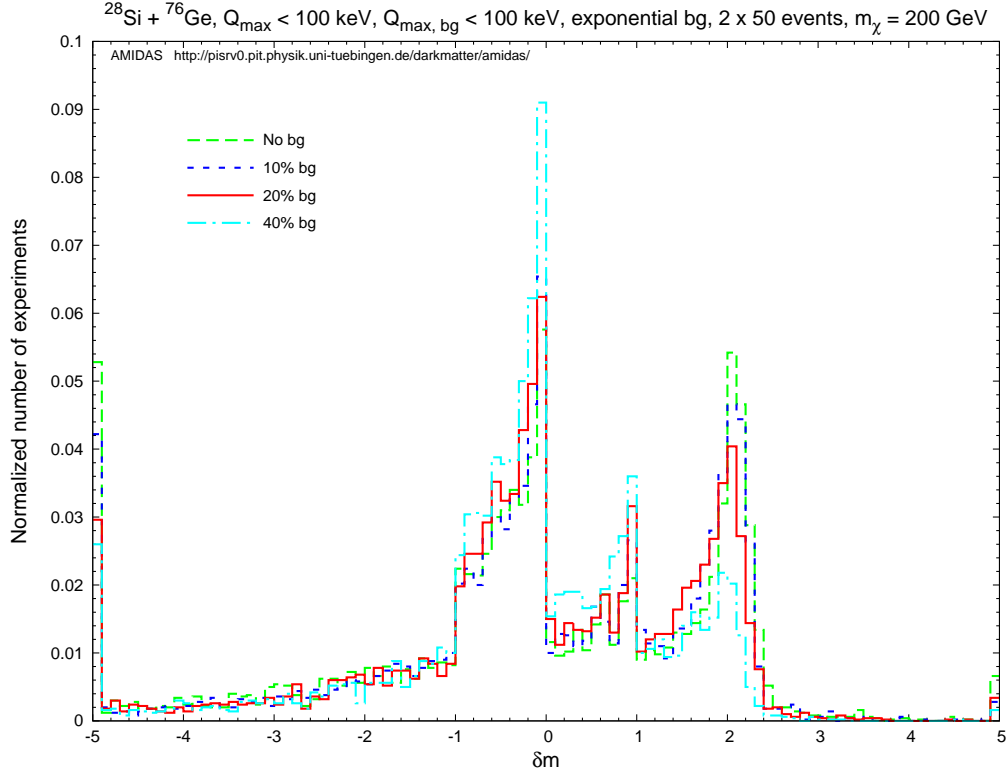}  \\ \vspace{1cm}
 \includegraphics[width=15cm]{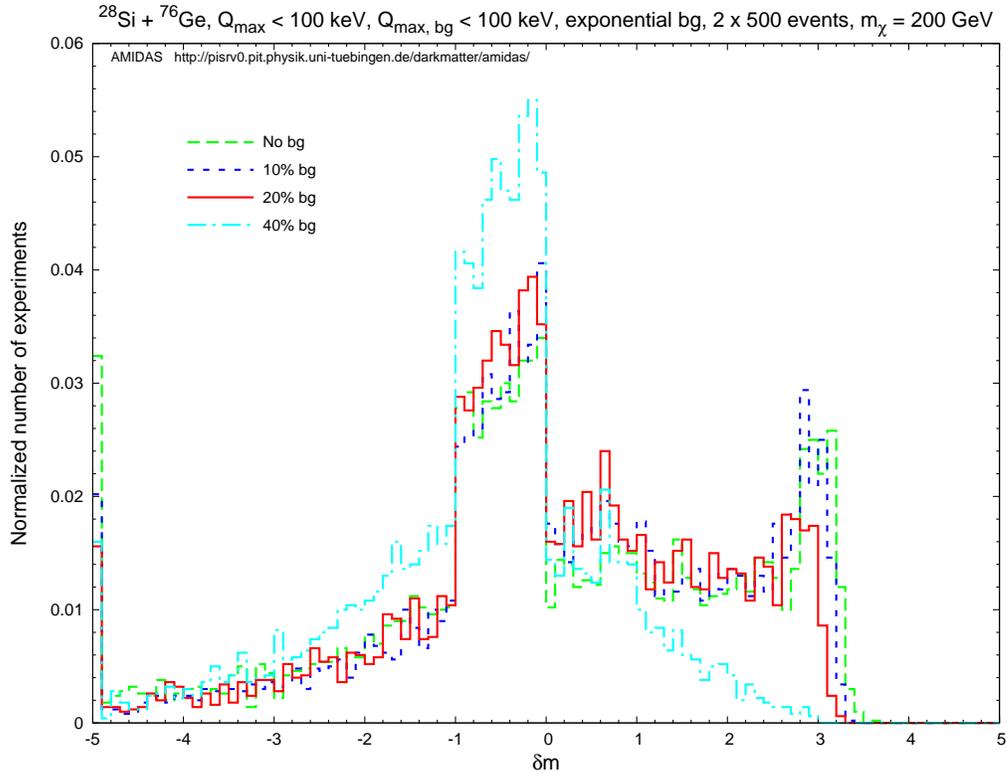}  \\
}
\vspace{-0.25cm}
\end{center}
\caption{
 Normalized distributions of the estimator $\delta m$
 defined in Eq.~(\ref{eqn:deltam})
 for an input WIMP mass of 200 GeV
 with 50 (upper) and 500 (lower) total events
 on average before cuts in each experiment.
 Parameters and notations are as
 in Fig.~\ref{fig:mchi-SiGe-ex-000-100-050}.
 Note that
 the bins at $\delta m = \pm 5$
 are overflow bins,
 i.e., they also contain all experiments with
 $|\delta m| > 5$.
 See the text for further details.
}
\label{fig:mchi-dev-SiGe-ex-000-100}
\end{figure}
 In Figs.~\ref{fig:mchi-dev-SiGe-ex-000-100}
 we show
 the normalized distributions of the estimator $\delta m$
 defined in Eq.~(\ref{eqn:deltam})
 for a rather heavy input WIMP mass of 200 GeV
 with 50 (upper) and 500 (lower) total events
 on average before cuts in each experiment.
 As discussed in Ref.~\cite{DMDDmchi},
 the deviation of the reconstructed WIMP mass
 in the simulated experiments
 looks asymmetric and non--Gaussian. 
 However,
 it can be seen here clearly that,
 the more the background events
 in our analyzed data sets,
 the more concentrated the $\delta m$ value
 in the range between $-1$ and 0
 as well as between 0 and $+1$. 
 Moreover,
 for the case with rather larger data sets
 of 500 total events,
 by increasing the background ratio
 the distribution becomes to be
 more symmetric and Gaussian--like,
 although the central value of $\delta m$
 seems to fall at $\sim -0.5$
 because of the underestimate of
 the reconstructed WIMP mass.

 In Ref.~\cite{DMDDmchi}
 it has been mentioned that
 with increasing number of total events
 the distribution of the estimator $\delta m$
 becomes slowly Gaussian.
 Figs.~\ref{fig:mchi-dev-SiGe-ex-000-100} here
 (and Fig.~\ref{fig:mchi-dev-SiGe-const-000-100-050} shown later also)
 indicate that
 with a {\em larger} background ratio
 in the analyzed data sets
 the distribution of $\delta m$
 approaches to be Gaussian {\em more fast}.
 This interesting observation
 might be able to offer some new ideas
 for improving the algorithmic procedure
 for the reconstruction of the WIMP mass
 with a higher statistical certainty. 
\subsubsection{With the constant background spectrum}
 In order to check the need of a prior knowledge about
 an (exact) form of the residue background spectrum,
 we consider briefly
 in this subsection
 a rather extrem case,
 i.e., the constant background spectrum
 in Eq.~(\ref{eqn:dRdQ_bg_const}).

\begin{figure}[b!]
\begin{center}
\imageswitch{
\begin{picture}(15,10.5)
\put(0,0){\framebox(15,10.5) {mchi-SiGe-const-000-100-050}}
\end{picture}}
{\includegraphics[width=15cm]{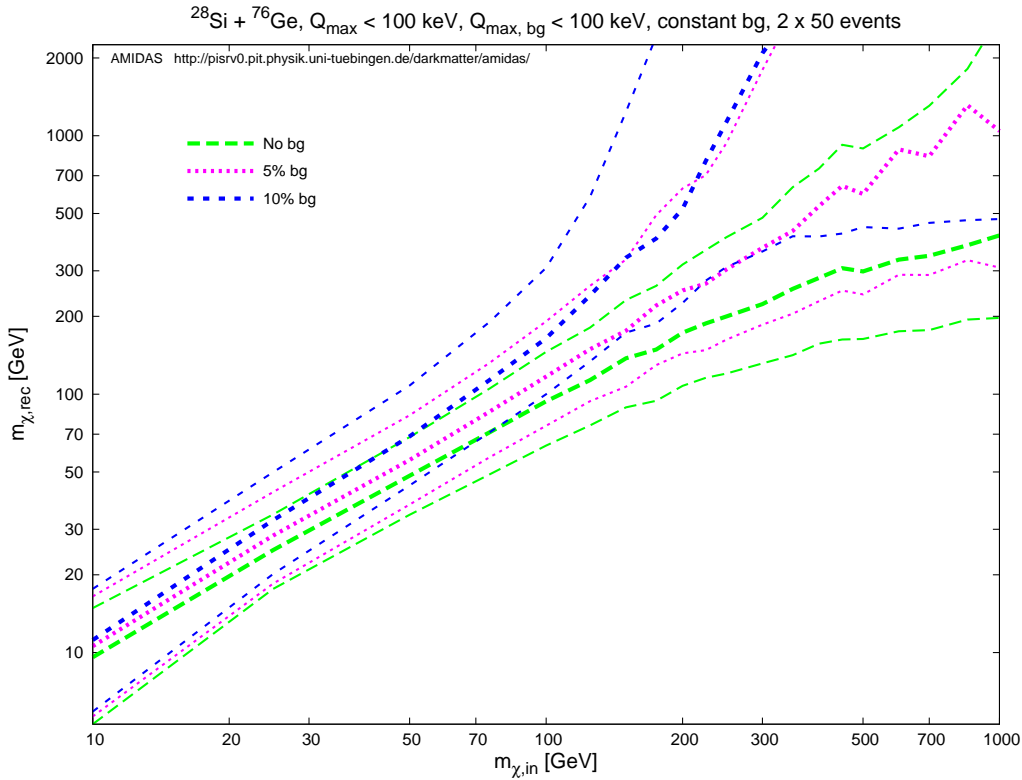} \\
}
\vspace{-0.5cm}
\end{center}
\caption{
 As in Fig.~\ref{fig:mchi-SiGe-ex-000-100-050},
 except that
 the constant background spectrum
 in Eq.~(\ref{eqn:dRdQ_bg_const})
 has been used.
 Note that
 the background ratios shown here
 are 5\% (dotted magenta curves)
 and 10\% (long--dotted blue curves).
}
\label{fig:mchi-SiGe-const-000-100-050}
\end{figure}

 In Fig.~\ref{fig:mchi-SiGe-const-000-100-050}
 we show the reconstructed WIMP mass
 and the lower and upper bounds of
 the 1$\sigma$ statistical uncertainty
 with mixed data sets
 as functions of the input WIMP mass.
 As in Figs.~\ref{fig:dRdQ-bg-const-Ge-000-100-20},
 the windows of the constant background spectrum
 are set as the same as
 the experimental possible energy ranges,
 i.e., between 0 and 100 keV
 for both experiments.
 The background ratios shown here
 are no background (dashed green curves),
  5\% (dotted magenta curves),
 and 10\% (long--dotted blue curves)
 background events in the whole data sets.
 Each experiment contains again 50 total events
 on average before cuts;
 all of these events are treated as WIMP signals
 in the analysis.

 It can be seen clearly that,
 as discussed above,
 since the constant background spectrum
 has relatively {\em flatter} shape
 compared to the WIMP scattering spectrum
 for not only light, but also heavy WIMP masses,
 and the measured energy spectrum should thus always
 mimic a scattering spectrum induced by {\em heavier} WIMPs,
 the reconstructed WIMP masses are therefore {\em overestimated}
 for all input WIMP masses,
 especially for the heavier masses.
 Actually,
 the result shown here looks more likely
 that shown in the lower frame of
 Figs.~\ref{fig:mchi-SiGe-ex-000-050-100-050},
 since in both cases
 residue background events
 contribute significantly more
 (compared to the exponential--like WIMP scattering spectrum)
 in high energy ranges.
 Not surprisingly,
 the larger the background ratio,
 the more strongly overestimated
 the reconstructed WIMP masses,
 in particular for the heavier input WIMP masses.
 Nevertheless,
 for (approximately) constant residue backgrounds
 with a fraction of $\sim$ 5\%
 in background windows
 as the entire experimental possible ranges,
 one could in principle still estimate the WIMP mass
 with a 1$\sigma$ statistical uncertainty as
 \mbox{$\sim 117~{\rm GeV}\~^{+64\%}_{-35\%}$} (for 100 GeV WIMPs) or
 $\sim  56~{\rm GeV}\~^{+49\%}_{-33\%}$ (for 50 GeV WIMPs).
 Once WIMPs are light ($\mchi \sim {\cal O}$(25 GeV)),
 the maximal acceptable background ratio
 could even be $\sim$ 10\%.

\begin{figure}[t!]
\begin{center}
\imageswitch{
\begin{picture}(15,10.5)
\put(0,0){\framebox(15,10.5) {mchi-SiGe-const-000-050-050}}
\end{picture}}
{\includegraphics[width=15cm]{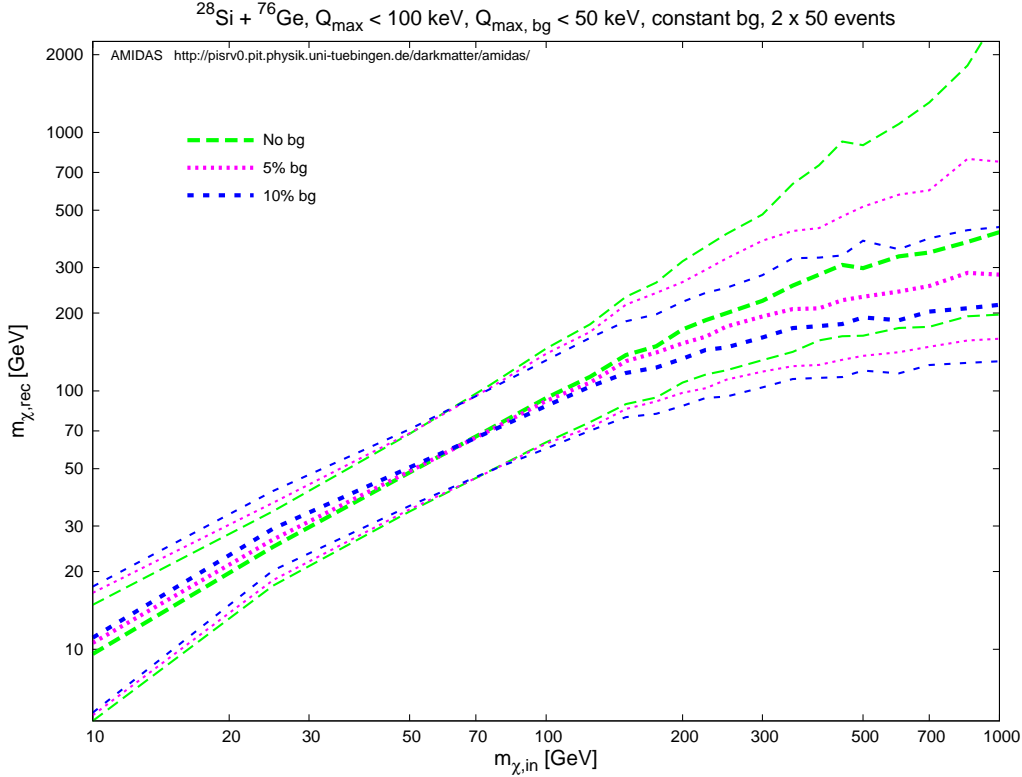} \\
}
\vspace{-0.5cm}
\end{center}
\caption{
 As in Fig.~\ref{fig:mchi-SiGe-const-000-100-050},
 except that
 the background window in each experiment
 has been set as 0 -- 50 keV.
}
\label{fig:mchi-SiGe-const-000-050-050}
\end{figure}
\begin{figure}[p!]
\begin{center}
\vspace{-0.75cm}
\imageswitch{
\begin{picture}(15,10.5)
\put(0,0){\framebox(15,10.5) {mchi-dev-SiGe-const-000-100-050}}
\end{picture}}
{\includegraphics[width=15cm]{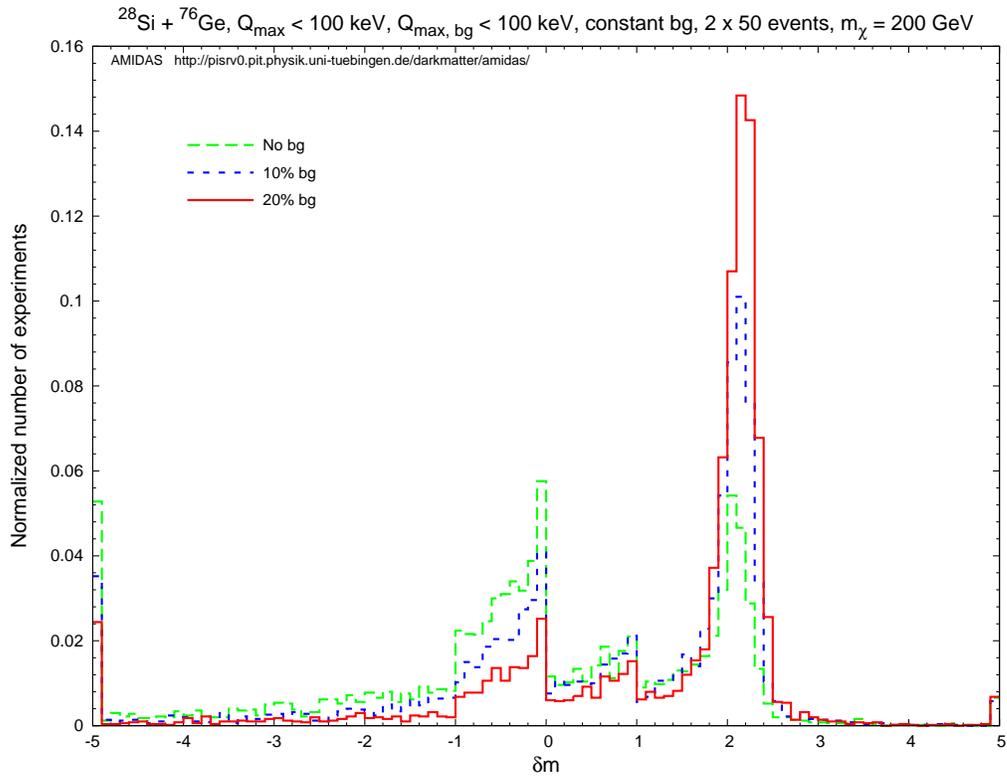} \\
}
\vspace{-0.5cm}
\end{center}
\caption{
 As in the upper frame of
 Figs.~\ref{fig:mchi-dev-SiGe-ex-000-100},
 except that
 the constant background spectrum
 in Eq.~(\ref{eqn:dRdQ_bg_const})
 has been used.
}
\label{fig:mchi-dev-SiGe-const-000-100-050}
\end{figure}
\begin{figure}[p!]
\begin{center}
\vspace{0.5cm}
\imageswitch{
\begin{picture}(15,10.5)
\put(0,0){\framebox(15,10.5) {mchi-SiGe-const-000-100-500}}
\end{picture}}
{\includegraphics[width=15cm]{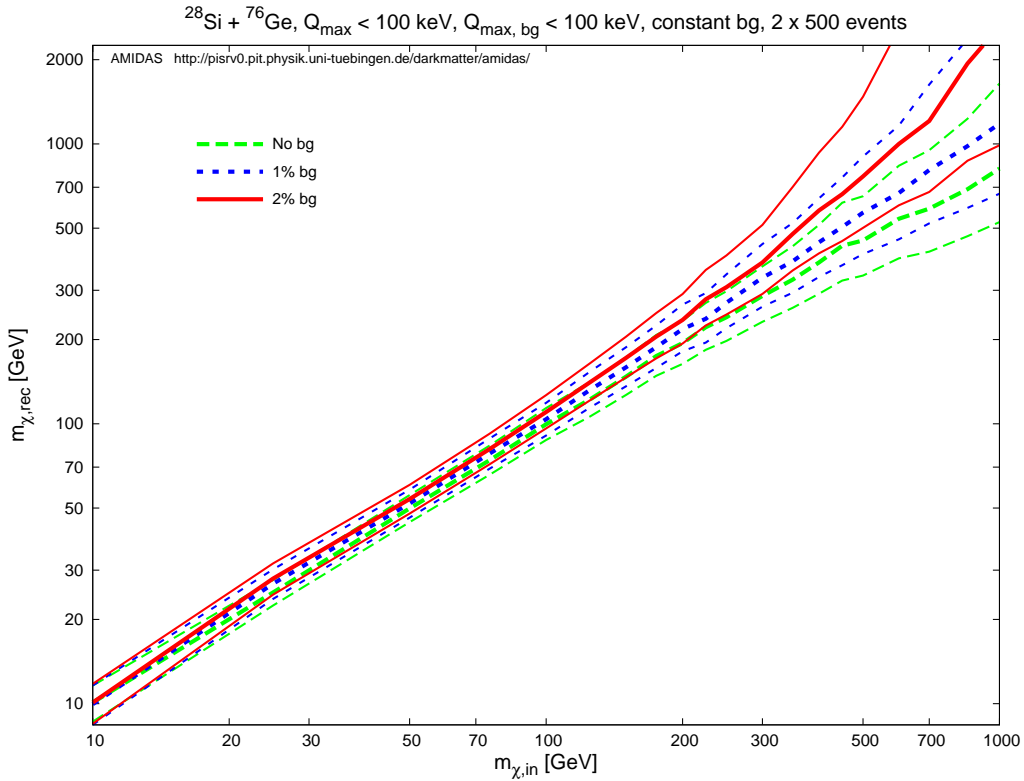} \\
}
\vspace{-0.5cm}
\end{center}
\caption{
 As in Fig.~\ref{fig:mchi-SiGe-const-000-100-050},
 except that
 the expected number of total events
 in both experiments
 has been set as 500.
 Note that
 the background ratios shown here
 are no background (dashed green curves),
% 0.5\% (dotted magenta curves),
 1\% (long--dotted blue curves),
 and 2\% (solid red curves),
 i.e., a factor of 10 smaller than
 the ratios used before.
 See the text for further details.
\vspace{-0.5cm}
}
\label{fig:mchi-SiGe-const-000-100-500}
\end{figure}

 Moreover,
 as done in Sec.~3.2.1,
 in Fig.~\ref{fig:mchi-SiGe-const-000-050-050}
 we shrink the background window in each experiment
 to the lower energy range between 0 and 50 keV.
 Not surprisingly,
 while for light WIMPs ($\mchi~\lsim~70$ GeV)%
\footnote{
 Remind that
 the actual value of this ``critical'' WIMP mass
 depends in practice strongly on the WIMP scattering spectrum
 as well as on the residue background spectrum
 and therefore differs from experiment to experiment.
},
 relatively more background events
 still contribute to high energy ranges;
 for heavy WIMPs ($\mchi~\gsim~70$ GeV),
 relatively more background events
 contribute now to low energy ranges and,
 consequently,
 the reconstructed WIMP masses are therefore {\em underestimated}
 for heavy WIMPs.

 On the other hand,
 as in Sec.~3.2.2,
 in Fig.~\ref{fig:mchi-dev-SiGe-const-000-100-050}
 we check
 the normalized distributions of the estimator $\delta m$
 for an input WIMP mass of 200 GeV
 with 50 total events
 on average before cuts in each experiment.
 It can be seen very clearly that,
 with increasing background ratio
 the value of $\delta m$ concentrates
 more and more strongly to 2.
 This means that,
 due to the contribution
 from residue background events,
 the reconstructed WIMP mass is most possibly
 $\sim$ 2$\sigma$ overestimated.
 Moreover,
 compared to the non--Gaussian form of the distributions
 for the case with the exponential background spectrum
 shown in the upper frame of
 Figs.~\ref{fig:mchi-dev-SiGe-ex-000-100},
 the distributions with the constant spectrum
 look more likely Gaussian,
 despite of the asymmetry
 due to the overestimate of the WIMP mass.
 Nevertheless,
 Figs.~\ref{fig:mchi-dev-SiGe-ex-000-100}
 and Fig.~\ref{fig:mchi-dev-SiGe-const-000-100-050}
 indicate that
 background events seem to let
 the distribution of the deviation of
 the reconstructed WIMP mass
 be more symmetric and Gaussian,
 no matter what kind of energy spectrum
 they would have.

 Finally,
 in Fig.~\ref{fig:mchi-SiGe-const-000-100-500}
 we rise the expected number of
 total events in each experiment
 by a factor of 10,
 to 500 events on average before cuts
 for the case that
 residue background events
 exist in the entire experimental possible energy ranges.
 Note that
 the background ratios shown here
 are no background (dashed green curves),
 1\% (long--dotted blue curves),
 and 2\% (solid red curves),
 i.e., a factor of 10 smaller than
 the ratios used before.
 In the lower frame of Figs.~\ref{fig:mchi-SiGe-ex-000-050-100-050}
 and in Fig.~\ref{fig:mchi-SiGe-const-000-100-050},
 we found that
 once $\sim$ 5\% -- 10\% events in our analyzed data sets
 are residue backgrounds and
 (most of) these events are recorded
 in high energy ranges,
 no matter what kind of spectrum shape
 they would have,
 the reconstructed WIMP mass
 could be (strongly) {\em overestimated}.
 However,
 Fig.~\ref{fig:mchi-SiGe-ex-000-100-500} and
 Fig.~\ref{fig:mchi-SiGe-const-000-100-500} here
 show that,
 by increasing the event number and
 decreasing the background ratio,
 one could in principle determine the WIMP mass
 (pretty) precisely
 {\em without} knowing the (exact) form of
 the spectrum of residue background events.
\section{Summary and conclusions}
 In this paper
 we reexamine the model--independent data analysis method
 introduced in Refs.~\cite{DMDDmchi-SUSY07, DMDDmchi}
 for the determination of the mass
 of Weakly Interacting Massive Particles
 from data (measured recoil energies) of
 direct Dark Matter detection experiments directly
 by taking into account a fraction of residue background events,
 which pass all discrimination criteria and
 then mix with other real WIMP--induced events
 in the analyzed data sets.
 Differ from the maximum likelihood analysis
 described in Refs.~\cite{Green-mchi07, Green-mchi08, Bernal08},
 our method requires {\em neither} prior knowledge
 about the WIMP scattering spectrum
 {\em nor} about different possible background spectra;
 the unique needed information is the recoil energies
 recorded in {\em two} direct detection experiments
 with {\em two different} target nuclei.

 In Sec.~2
 we considered first the measured energy spectrum
 for different WIMP masses
 with two forms of possible residue background spectrum:
 the {\em target--dependent exponential} spectrum
 and the {\em constant} spectrum.
 The exponential background spectrum
 contributes relatively more events
 to {\em high} energy ranges
 once WIMPs are {\em light} ($\mchi~\lsim~100$ GeV),
 and to {\em low} energy ranges
 for {\em heavy} WIMP masses ($\mchi~\gsim~100$ GeV);
 whereas
 the constant background spectrum
 contributes always relatively more events
 to {\em high} energy ranges.
 As the consequence,
 the energy spectrum of all observed events
 looks more likely to be a scattering spectrum
 induced by {\em heavier} WIMPs,
 once the spectrum of residue background events
 (induced perhaps by two or more different sources)
 is either exponential--like (and WIMPs are light)
 or approximately constant
 (for all WIMP masses);
 while
 if WIMPs are heavy and
 the residue background spectrum
 is approximately exponential,
 the measured energy spectrum
 would look more likely to be a scattering spectrum
 induced by {\em lighter} WIMPs.

 In Sec.~3.2
 the data sets generated in Sec.~2
 have been analyzed for reconstructing
 the mass of incident WIMPs
 by using the model--independent method.
 With the exponential background spectrum,
 the input WIMP mass would be {\em overestimated}
 once WIMPs are light ($\mchi~\lsim~100$ GeV),
 or, in contrast,
 would be {\em underestimated}
 for heavy WIMPs ($\mchi~\gsim~100$ GeV).
 Our simulations show that,
 for background windows in the {\em entire or low}
 experimental possible energy ranges,
 one could in principle reconstruct the WIMP mass
 with a maximal fraction of $\sim$ 20\%
 of residue background events in the analyzed data sets;
 whereas
 for background windows in {\em high} energy ranges,
 the maximal acceptable fraction
 of residue backgrounds
 is only $\sim$ 10\%.

 Moreover,
 in order to check the need of a prior knowledge about
 an (exact) form of the residue background spectrum,
 we considered also the case
 with the constant background spectrum.
 In this rather extrem case,
 the WIMP mass would always be {\em overestimated},
 especially for heavy WIMPs ($\mchi~\gsim~100$ GeV).
 Our simulations give then
 a maximal acceptable fraction of \mbox{$\sim$ 5\% -- 10\%}
 of residue background events in the data sets
 for background windows in the {\em entire or low}
 experimental possible energy ranges.
 Nevertheless,
 we found also that,
 by means of
 increased number of observed (WIMP--induced) events
 and improved background discrimination techniques
 \cite{CRESST-bg, % Lang09a, Schmaler09,
       EDELWEISS-bg}, % Broniatowski09, Armengaud09}
 the WIMP mass could in principle be determined
 (pretty) precisely,
 no matter what kind of energy spectrum
 residue background events would have.

 On the other hand,
 in order to check the statistical fluctuation
 of the reconstructed WIMP mass
 with increased background ratio,
 we considered also the distribution of
 the deviation of the reconstructed WIMP mass
 from the true value.
 It was found in Ref.~\cite{DMDDmchi} that,
 for a rather heavy WIMP mass of 200 GeV,
 the distribution of the deviation of
 the reconstructed WIMP mass
 is asymmetric and non--Gaussian,
 either with data sets of only a few ($\cal O$(50)) events
 or with larger date sets (of $\cal O$(500) events).
 However,
 our simulations with different background ratios
 show that,
 firstly,
 for both used (exponential and constant) background spectra,
 with increasing background ratio
 the distribution of the deviation of
 the reconstructed WIMP mass
 becomes more and more concentrated,
 although still asymmetric and non--Gaussian.
 Secondly,
 for the more realistic
 exponential background spectrum
 and using data sets with a larger number of total events,
 with increasing background ratio
 the distribution of the deviation
 becomes somehow more symmetric and Gaussian.
 This observation might be able to offer some new ideas
 for improving the algorithmic procedure
 for the reconstruction of the WIMP mass
 with a higher statistical certainty.

 In summary,
 our study of the effects of residue background events
 in direct Dark Matter detection experiments
 on the determination of the WIMP mass
 shows that,
 with currently running and projected experiments
 using detectors with $10^{-9}$ to $10^{-11}$ pb sensitivities
 \cite{Baudis07a, Drees08, Aprile09a, Gascon09}
 and $< 10^{-6}$ background rejection ability
 \cite{CRESST-bg, % Lang09a,
       EDELWEISS-bg, % Broniatowski09,
       Lang09b, % Armengaud09,
       Ahmed09b}, % , Schmaler09},
 once two or more experiments with different target nuclei
 could accumulate a few tens events
 (in one experiment),
 we could in principle already estimate
 the mass of Dark Matter particle
 with a reasonable precision,
 even though there might be some background events
 mixed in our data sets for the analysis%
\footnote{
 A possible first test could be a combination of
 the events observed
 by the CoGeNT experiment with their Ge detector
 with the events observed
 in the oxygen band of the CRESST-II experiment
 \cite{Aalseth10, CRESST-Talk, Hooper10}.
}.
 Moreover,
 two forms for background spectrum
 and three windows for residue background events
 considered in this work are rather naive.
 Nevertheless,
 one should be able to extend our observations/discussions
 to predict the effects of possible background events
 in their own experiment.
 Hopefully,
 this will encourage our experimental colleagues
 to present their (future) results
 not only in form of the ``exclusion limit(s)'',
 but also of the ``most possible area(s)''
 on the cross section versus mass plan.
\subsubsection*{Acknowledgments}
 The authors would like to thank
 the Physikalisches Institut der Universit\"at T\"ubingen
 for the technical support of the computational work
 demonstrated in this article.
 CLS would also like to thank
 the friendly hospitality of the
 Max--Planck--Institut f\"ur Kernphysik in Heidelberg
 where part of this work was completed.
 This work
 was partially supported by
 the National Science Council of R.O.C.~%
 under contracts no.~NSC-96-2112-N-009-023-MY3 and
 no.~NSC-98-2811-M-006-044
 as well as by
 the LHC Physics Focus Group,
 National Center of Theoretical Sciences, R.O.C..
\appendix
\setcounter{equation}{0}
\setcounter{figure}{0}
\renewcommand{\theequation}{A\arabic{equation}}
\renewcommand{\thefigure}{A\arabic{figure}}
%
%
% Appendix A
\section{Formulae needed in Sec.~3}
 Here we list all formulae needed
 for the model--independent data analyses procedure
 used in Sec.~3.
 Detailed derivations and discussions
 can be found in Refs.~\cite{DMDDf1v, DMDDmchi}.
\subsection{Estimating \boldmath$r(\Qmin)$, $I_n(\Qmin, \Qmax)$,
            and their statistical errors}
 Firstly,
 consider experimental data described by
\beq
     {\T Q_n - \frac{b_n}{2}}
 \le \Qni
 \le {\T Q_n + \frac{b_n}{2}}
\~,
     ~~~~~~~~~~~~ %12
     i
 =   1,~2,~\cdots,~N_n,~
     n
 =   1,~2,~\cdots,~B.
\label{eqn:Qni}
\eeq
 Here the entire experimental possible
 energy range between $\Qmin$ and $\Qmax$
 has been divided into $B$ bins
 with central points $Q_n$ and widths $b_n$.
 In each bin,
 $N_n$ events will be recorded.
 Since the recoil spectrum $dR / dQ$ is expected
 to be approximately exponential,
 the following ansatz for the {\em measured} recoil spectrum
 ({\em before} normalized by the exposure $\calE$)
 in the $n$th bin has been introduced \cite{DMDDf1v}:
\beq
        \adRdQ_{{\rm expt}, \~ n}
 \equiv \adRdQ_{{\rm expt}, \~ Q \simeq Q_n}
 \equiv \rn  \~ e^{k_n (Q - Q_{s, n})}
\~.
\label{eqn:dRdQn}
\eeq
 Here $r_n$ is the standard estimator
 for $(dR / dQ)_{\rm expt}$ at $Q = Q_n$:
\beq
   r_n
 = \frac{N_n}{b_n}
\~,
\label{eqn:rn}
\eeq
 $k_n$ is the logarithmic slope of
 the recoil spectrum in the $n$th $Q-$bin,
 which can be computed numerically
 from the average value of the measured recoil energies
 in this bin:
\beq
   \bQn
 = \afrac{b_n}{2} \coth\afrac{k_n b_n}{2}-\frac{1}{k_n}
\~,
\label{eqn:bQn}
\eeq
 where
\beq
        \bQxn{\lambda}
 \equiv \frac{1}{N_n} \sumiNn \abrac{\Qni - Q_n}^{\lambda}
\~.
\label{eqn:bQn_lambda}
\eeq
 The error on the logarithmic slope $k_n$
 can be estimated from Eq.~(\ref{eqn:bQn}) directly as
\beq
   \sigma^2(k_n)
 = k_n^4
   \cbrac{  1
          - \bfrac{k_n b_n / 2}{\sinh (k_n b_n / 2)}^2}^{-2}
            \sigma^2\abrac{\bQn}
\~,
\label{eqn:sigma_kn}
\eeq
 with
\beq
   \sigma^2\abrac{\bQn}
 = \frac{1}{N_n - 1} \bbigg{\bQQn - \bQn^2}
\~.
\label{eqn:sigma_bQn}
\eeq
 $Q_{s, n}$ in the ansatz (\ref{eqn:dRdQn})
 is the shifted point at which
 the leading systematic error due to the ansatz
 is minimal \cite{DMDDf1v},
\beq
   Q_{s, n}
 = Q_n + \frac{1}{k_n} \ln\bfrac{\sinh(k_n b_n/2)}{k_n b_n/2}
\~.
\label{eqn:Qsn}
\eeq
 Note that $Q_{s, n}$ differs from
 the central point of the $n$th bin, $Q_n$.
 From the ansatz (\ref{eqn:dRdQn}),
 the counting rate at $Q = \Qmin$ can be calculated by
\beq
   r(\Qmin)
 = r_1 e^{k_1 (\Qmin - Q_{s, 1})}
\~,
\label{eqn:rmin_eq}
\eeq
 and its statistical error can be expressed as
\beq
   \sigma^2(r(\Qmin))
 = r^2(\Qmin) 
   \cbrac{  \frac{1}{N_1}
          + \bbrac{  \frac{1}{k_1}
                   - \afrac{b_1}{2} 
                     \abrac{  1
                            + \coth\afrac{b_1 k_1}{2}}}^2
            \sigma^2(k_1)}
\~,
\label{eqn:sigma_rmin}
\eeq
 since
\beq
   \sigma^2(r_n)
 = \frac{N_n}{b_n^2}
\~.
\label{eqn:sigma_rn}
\eeq
 Finally,
 since all $I_n$ are determined from the same data,
 they are correlated with
\beq
   {\rm cov}(I_n, I_m)
 = \sum_a \frac{Q_a^{(n+m-2)/2}}{F^4(Q_a)}
\~,
\label{eqn:cov_In}
\eeq
 where the sum runs again over all events
 with recoil energy between $\Qmin$ and $\Qmax$. 
 And the correlation between the errors on $r(\Qmin)$,
 which is calculated entirely
 from the events in the first bin,
 and on $I_n$ is given by
\beqn
 \conti {\rm cov}(r(\Qmin), I_n)
        \non\\
 \=     r(\Qmin) \~ I_n(\Qmin, \Qmin + b_1)
        \non\\
 \conti ~~~~ \times %4
        \cleft{  \frac{1}{N_1} 
               + \bbrac{  \frac{1}{k_1}
                        - \afrac{b_1}{2} \abrac{1 + \coth\afrac{b_1 k_1}{2}}}}
        \non\\
 \conti ~~~~~~~~~~~~~~ \times % 14
        \cright{ \bbrac{  \frac{I_{n+2}(\Qmin, \Qmin + b_1)}
                               {I_{n  }(\Qmin, \Qmin + b_1)}
                        - Q_1
                        + \frac{1}{k_1}
                        - \afrac{b_1}{2} \coth\afrac{b_1 k_1} {2}}
        \sigma^2(k_1)}
\~;
\label{eqn:cov_rmin_In}
\eeqn
 note that
 the sums $I_i$ here only count in the first bin,
 which ends at $Q = \Qmin + b_1$.
\subsection{Statistical errors on \boldmath$\mchi$
            given in Eqs.~(\ref{eqn:mchi_Rn}) and (\ref{eqn:mchi_Rsigma})}
 By using the standard Gaussian error propagation,
 a lengthy expression for the statistical error on
 $\left. \mchi \right|_{\Expv{v^n}}$
 given in Eq.~(\ref{eqn:mchi_Rn})
 can be obtained as
\beqn
        \left. \sigma(\mchi) \right|_{\Expv{v^n}}
 \=     \frac{\sqrt{\mX / \mY} \vbrac{\mX - \mY} \abrac{\calR_{n, X} / \calR_{n, Y}} }
             {\abrac{\calR_{n, X} / \calR_{n, Y} - \sqrt{\mX / \mY}}^2}
        \non\\
 \conti ~~~~ \times
        \bbrac{  \frac{1}{\calR_{n, X}^2}
                 \sum_{i, j = 1}^3
                 \aPp{\calR_{n, X}}{c_{i, X}} \aPp{\calR_{n, X}}{c_{j, X}}
                 {\rm cov}(c_{i, X}, c_{j, X})
               + (X \lto Y)}^{1 / 2}
\!.
\label{eqn:sigma_mchi_Rn}
\eeqn
 Here a short--hand notation for the six quantities
 on which the estimate of $\mchi$ depends has been introduced:
\beq
   c_{1, X}
 = I_{n, X}
\~,
   ~~~~~~~~~~~~ %12
   c_{2, X}
 = I_{0, X}
\~,
   ~~~~~~~~~~~~ %12
   c_{3, X}
 = r_X(\QminX)
\~;
\label{eqn:ciX}
\eeq
 and similarly for the $c_{i, Y}$.
 Estimators for ${\rm cov}(c_i, c_j)$ have been given
 in Eqs.~(\ref{eqn:cov_In}) and (\ref{eqn:cov_rmin_In}).
 Explicit expressions for the derivatives of $\calR_{n, X}$
 with respect to $c_{i, X}$ are:
\cheqnXa{A}
\beq
   \Pp{\calR_{n, X}}{\InX}
 = \frac{n + 1}{n}
   \bfrac{\FQminX}{2 \QminX^{(n + 1) / 2} r_X(\QminX) + (n + 1) \InX \FQminX}
   \calR_{n, X}
\~,
\label{eqn:dRnX_dInX}
\eeq
\cheqnXb{A}
\beq
   \Pp{\calR_{n, X}}{\IzX}
 =-\frac{1}{n}
   \bfrac{\FQminX}{2 \QminX^{1 / 2} r_X(\QminX) + \IzX \FQminX}
   \calR_{n, X}
\~,
\label{eqn:dRnX_dIzX}
\eeq
 and
\cheqnXc{A}
\beqn
        \Pp{\calR_{n, X}}{r_X(\QminX)}
 \=     \frac{2}{n}
        \bfrac{  \QminX^{(n + 1) / 2} \IzX        - (n + 1) \QminX^{1 / 2} \InX}
              {2 \QminX^{(n + 1) / 2} r_X(\QminX) + (n + 1) \InX \FQminX}
        \non\\
 \conti ~~~~~~~~~~~~~~~~ \times %16
        \bfrac{\FQminX}{2 \QminX^{1 / 2} r_X(\QminX) + \IzX \FQminX}
        \calR_{n, X}
\~;
\label{eqn:dRnX_drminX}
\eeqn
\cheqnX{A}%
 explicit expressions for the derivatives of $\calR_{n, Y}$
 with respect to $c_{i, Y}$ can be given analogously.
 Note that,
 firstly,
 factors $\calR_{n, (X, Y)}$ appear in all these expressions,
 which can practically be cancelled by the prefactors
 in the bracket in Eq.~(\ref{eqn:sigma_mchi_Rn}).
 Secondly,
 all the $I_{0, (X, Y)}$ and $I_{n, (X, Y)}$ should be understood
 to be computed according to
 Eq.~(\ref{eqn:In_sum}) % or (\ref{eqn:In_int})
 with integration limits $\Qmin$ and $\Qmax$
 specific for that target.

 Similar to the analogy between
 Eqs.~(\ref{eqn:mchi_Rn}) and (\ref{eqn:mchi_Rsigma}),
 the statistical error on $\left. \mchi \right|_\sigma$
 given in Eq.~(\ref{eqn:mchi_Rsigma})
 can be expressed as
\beqn
        \left. \sigma(\mchi) \right|_\sigma
 \=     \frac{\abrac{\mX / \mY}^{5 / 2} \vbrac{\mX - \mY}
              \abrac{\calR_{\sigma, X} / \calR_{\sigma, Y}} }
             {\bbrac{\calR_{\sigma, X} / \calR_{\sigma, Y} - \abrac{\mX / \mY}^{5 / 2}}^2}
        \non\\
 \conti ~~~~~~ \times %6
        \bbrac{  \frac{1}{\calR_{\sigma, X}^2}
                 \sum_{i, j = 2}^3
                 \aPp{\calR_{\sigma, X}}{c_{i, X}} \aPp{\calR_{\sigma, X}}{c_{j, X}}
                 {\rm cov}(c_{i, X}, c_{j, X})
               + (X \lto Y)}^{1 / 2}
\~,
\label{eqn:sigma_mchi_Rsigma}
\eeqn
 where we have again used
 the short--hand notation in Eq.~(\ref{eqn:ciX});
 note that $c_{1, (X, Y)} = I_{n, (X, Y)}$ does not appear here.
 Expressions for the derivatives of $\calR_{\sigma, X}$
 can be computed from Eq.~(\ref{eqn:RsigmaX_min}) as
\cheqnXa{A}
\beq
   \Pp{\calR_{\sigma, X}}{\IzX}
 = \bfrac{\FQminX}{2 \QminX^{1 / 2} r_X(\QminX) + \IzX \FQminX}
   \calR_{\sigma, X}
\~,
\label{eqn:dRsigmaX_dIzX}
\eeq
\cheqnXb{A}
\beq
   \Pp{\calR_{\sigma, X}}{r_X(\QminX)}
 = \bfrac{2 \QminX^{1 / 2}}{2 \QminX^{1 / 2} r_X(\QminX) + \IzX \FQminX}
   \calR_{\sigma, X}
\~;
\label{eqn:dRsigmaX_drminX}
\eeq
\cheqnX{A}%
 and similarly for the derivatives of $\calR_{\sigma, Y}$.
\subsection{Covariance of \boldmath$f_i$
            defined in Eqs.~(\ref{eqn:fiXa}) and (\ref{eqn:fiXb})}
 The entries of the $\cal C$ matrix
 given in Eq.~(\ref{eqn:Cij})
 involving basically only
 the moments of the WIMP velocity distribution
 can be read off Eq.~(82) of Ref.~\cite{DMDDf1v},
 with an slight modification
 due to the normalization factor in Eq.~(\ref{eqn:fiXa})%
\footnote{
 Since the last $f_i$ defined in Eq.~(\ref{eqn:fiXb})
 can be computed from the same basic quantities,
 i.e., the counting rates at $\Qmin$ and the integrals $I_0$,
 it can directly be included in the covariance matrix.
}:
\beqn
        {\rm cov}\abrac{f_i, f_j}
 \=     \calN_{\rm m}^2
        \bbiggl{  f_i \~ f_j \~ {\rm cov}(I_0, I_0)
                + \Td{\alpha}^{i+j} (i+1) (j+1) {\rm cov}(I_i, I_j)}
        \non\\
 \conti ~~~~~~~~ %8
                - \Td{\alpha}^j (j+1) f_i \~ {\rm cov}(I_0, I_j)
                - \Td{\alpha}^i (i+1) f_j \~ {\rm cov}(I_0, I_i)\bigg.
        \non\\
 \conti ~~~~~~~~~~~~ %12
                + D_i D_j \sigma^2(r(\Qmin))
                - \abrac{D_i f_j + D_j f_i} {\rm cov}(r(\Qmin), I_0)\Bigg.
        \non\\
 \conti ~~~~~~~~~~~~~~~~ %16
        \bbiggr{+ \Td{\alpha}^j (j+1) D_i \~ {\rm cov}(r(\Qmin), I_j)
                + \Td{\alpha}^i (i+1) D_j \~ {\rm cov}(r(\Qmin), I_i)}
\~.
        \non\\
\label{eqn:cov_fi}
\eeqn
 Here we have defined
\beq
        \calN_{\rm m}
 \equiv \frac{1}{2 \Qmin^{1 /2} r(\Qmin)/\FQmin + I_0}
\~,
\label{eqn:calNm}
\eeq
\beq
        \Td{\alpha}
 \equiv \frac{\alpha}{300~{\rm km/s}}
\~,
\label{eqn:td_alpha}
\eeq
 and
\cheqnXa{A}
\beq
        D_i
 \equiv \frac{1}{\cal N_{\rm m}} \bPp{f_i}{r(\Qmin)}
 =      \frac{2}{\FQmin}
        \abigg{\Td{\alpha}^i \Qmin^{(i+1)/2} - \Qmin^{1/2} \~ f_i}
\~,
\label{eqn:Dia}
\eeq
 for $i = -1,~1,~2,~\dots,~n_{\rm max}$; and
\cheqnXb{A}
\beq
   D_{n_{\rm max}+1}
 = \frac{2}{\FQmin} \abrac{-\Qmin^{1/2} f_{n_{\rm max}+1}}
\~.
\label{eqn:Dib}
\eeq
\cheqnX{A}%
\end{document}